\documentclass[prb,twocolumn]{revtex4}
\usepackage{graphics}

\newcommand{\be}{\begin{equation}}
\newcommand{\ee}{\end{equation}}
\newcommand{\bea}{\begin{eqnarray}}
\newcommand{\eea}{\end{eqnarray}}

\begin{document}

\title{Collective Modes of $\nu =2$ Quantum Hall Bilayers in Tilted Magnetic Field}
\author{Anna Lopatnikova$^{1*}$}
\author{Steven H. Simon$^{2}$}
\author{Eugene Demler$^{1}$}
\affiliation{
$^{1}$Department of Physics, Harvard University, Cambridge, MA 02138 \\
$^{2}$Lucent Technologies Bell Labs, Murray Hill, NJ 07974 }

\date{\today}

\begin{abstract}

We use time-dependent Hartree Fock approximation to study the
collective mode spectra of $\nu=2$ quantum Hall bilayers in tilted
magnetic field allowing for charge imbalance as well as tunneling
between the two layers.  In a previous companion paper to this
work, we studied the zero temperature global phase diagram of this
system which was found to include a symmetric and ferromagnetic
phases as well as a first order transition between two canted
phases with spontaneously broken $U(1)$ symmetry. We further found
that this first order transition line ends in a quantum critical
point within the canted region.  In the current work, we study the
excitation spectra of all of these phases and pay particular
attention to the behavior of the collective modes near the phase
transitions.  We find, most interestingly, that the first order
transition between the two canted phases is signaled by a near
softening of a magnetoroton minimum.  Many of the collective mode
features explored here should be accessible experimentally in
light scattering experiments.

\end{abstract}

\maketitle
\section{Introduction}

Scattering experiments have provided extremely powerful and
important probes of two dimensional electron
systems\cite{dassarma}.   A particularly nice application of light
scattering was a recent set
experiments\cite{pellegrini:97,pellegrini:98} on quantum Hall
bilayers with equal densities in each layer. An apparent mode
softening at total filling fraction $\nu=2$ was identified with
the existence of a Goldstone mode which fit well with prior
predictions of a novel canted phase in these bilayer
systems\cite{zheng:97,dassarma:98}. The experiments were conducted
using the tilted-field technique for sweeping across a wide range
of Zeeman energies {\it in situ}. Interestingly, tilted magnetic
fields also have a non-trivial effect on interlayer tunneling.
Furthermore,  as noted first by Burkov and MacDonald
\cite{burkov:02}, in $\nu=2$ bilayer systems with a {\it
charge-imbalance} between the layers, tilting the magnetic field
can induce a new first order quantum phase transition embedded in
the canted phase. As shown by the current authors in the preceding
companion paper to this work\cite{lopatnikova:03a}, the first
order transition separates two phases with the same symmetry which
are topologically connected in the phase diagram.  Similar to a
liquid-gas transition, the first order phase transition line
terminates at a quantum critical point.  In our preceding
paper\cite{lopatnikova:03a} we discussed the phases and phase
transitions of $\nu=2$ in detail, accounting for both charge
imbalance and in-plane magnetic field. The purpose of the present
work is to examine the excitation spectra of these different
phases in order to make connection with possible future
experiments.   Particular attention will be paid to the evolution
of the collective-mode dispersions across the new first-order
transition induced by the tilted magnetic field.

Bilayer quantum Hall systems in general have been the focus of a
great deal of recent study\cite{dassarma2}.  The already rich
physics of quantum Hall effects is further enhanced in bilayers by
the added degree of freedom.  The most studied of the bilayer
quantum Hall state is certainly the $\nu=1$ state\cite{dassarma2}.
At $\nu=1$ the spin degrees of freedom are effectively frozen out,
and all the interesting physics occurs in the isospin (layer
degrees of freedom).  In contrast, for $\nu=2$ systems, not only
the layer but also the spin degrees of freedom are important.
Indeed, in several of the phases of $\nu=2$, the spin and isospin
degrees of freedom are actually entangled.

In perpendicular magnetic field, $\nu=2$ bilayers exhibit the
many-body canted phase at finite tunneling\cite{zheng:97}.  This
phase is a spontaneously-broken $U(1)$ symmetry phase, despite the
finite tunneling. This is in marked contrast with the $\nu=1$
bilayers, in which a $U(1)$ symmetric phase is possible only in
the absence of tunneling. Things change when a finite voltage bias
is added: In this case, $\nu=2$ bilayers can exhibit a many-body
phase in the absence of tunneling.  This phase is somewhat akin to
the many-body phase of $\nu=1$ bilayers, as was pointed out by
MacDonald, Rajaraman, and Jungwirth\cite{macdonald:99}. These
authors therefore mused that, in the presence of a finite in-plane
magnetic field component, $\nu=2$ bilayers may also undergo a
commensurate incommensurate transition.

\begin{figure}[b]
        \centering
        \scalebox{0.75}{\includegraphics{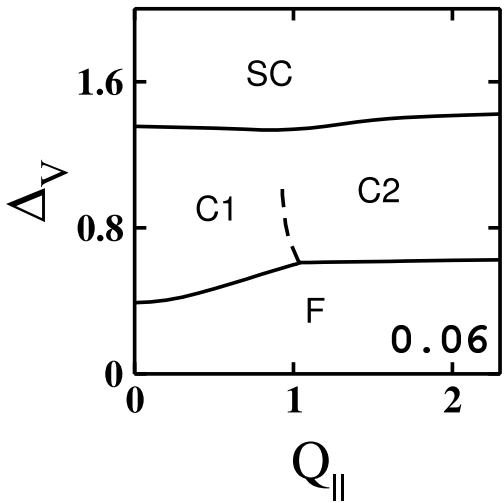}}
    \caption[Phase diagram of charge-unbalanced $\nu=2$ bilayers
     in tilted magnetic field.]{\label{fig:c1c2single} Global phase
      diagram of charge-unbalanced $\nu=2$ bilayers in tilted magnetic field.
      The phase diagram is calculated for bare tunneling gap $\Delta_{SAS}^0 =
       0.06 e^2/(\varepsilon l)$, Zeeman energy
       $\Delta_Z^0 = 0.01 e^2/(\varepsilon l)$,
       and the distance between the layers $d = l$.  The axes are
       the amplitude of the in-plane field wavevector: $\vec{Q}_{||} =
       \frac{\hat{z}\times \vec{B}_{||}}{(B_{\perp }l^2/d)}$ and the
       external bias voltage $\Delta_V$.  This choice of axes is particularly
        suitable, since current experimental techniques allow to vary the
         bias voltage and the in-plane field {\it in situ} over a wide
          range of values. $SC$ is the spin-singlet commensurate phase,
           $C1$ is the simple, isospin-commensurate phase, $C2$ is the
            spin-isospin commensurate phase.  Solid lines represent
             the second-order quantum phase transitions, dashed line
             --- the novel first-order $C1$--$C2$ transition.}
\end{figure}

Burkov and MacDonald\cite{burkov:02} explored this possibility.
Indeed, they found that {\it charge-unbalanced} $\nu=2$ bilayers
can undergo a phase transition driven by the in-plane field
component. However, the phase transition was between two
commensurate phases, instead of a between a commensurate and an
incommensurate phase. One of the commensurate phases was akin to
the commensurate phase of the $\nu=1$ bilayers -- the isospin
component followed the magnetic field.  The other commensurate
phase however, was more peculiar: in this phase, both isospin and
spin components were commensurate with the in-plane field! In our
previous publication, we attempted to understand the physics
behind this spin-commensuration.  We explored the phase transition
further, and found that it terminates at a critical point within
the canted phase (See Fig.~\ref{fig:c1c2single}).

As mentioned above, one way of experimentally distinguishing
between the many phases of the $\nu=2$ bilayers is to probe the
collective excitations \cite{pellegrini:97,pellegrini:98}.   In
this paper we therefore set out to explore this phase transition
further by finding the collective modes.  The many-body phases
($C1$, and $C2$ as well as $I$ which occurs in the absence of
tunneling) of the $\nu=2$ bilayers are characterized by
spontaneously broken symmetries, which result in the formation of
low-energy Goldstone modes.  In fact, the first theoretical and
experimental evidence of the canted phase in charge-balanced
$\nu=2$ bilayers was obtained by observing a softening
spin-density mode in time-dependent Hartree-Fock (TDHF) analysis
\cite{zheng:97} and in inelastic light-scattering experiments
\cite{pellegrini:97,pellegrini:98}.

More generally, the TDHF approximation allows one to predict the
response of the system to any one of a number of possible
experimental probes \cite{kallin:84,cote:91}.  A general
perturbation can be introduced into the system by the addition of
a small time-dependent term, $\delta H$: \be \label{eq:pertH1}
    \delta H = \sum_{\mu \sigma \nu \sigma'} \int \frac{d\omega}{2 \pi} \int \frac{d^2 k}{2
    \pi}\,\, \,
    \Phi_{\mu \sigma ,\nu \sigma'}({\bf k}, \omega)  e^{-i \omega t}\, \, \rho_{\mu \sigma ,\nu \sigma'}^\dagger({\bf k}, \omega)
\ee to the total Hamiltonian (Eq.~(\ref{eq:H})).  The operators
$\rho_{\mu \sigma, \nu \sigma'}$ are the density operators \be
    \rho_{\mu\sigma ,\nu\sigma' }({\bf k},0) =
    \frac{1}{g }\sum_{X} e^{-ik_xX-ik_xk_yl^2/2}c^{\dag }_{\mu \sigma ,X}c_{\nu \sigma' ,X+k_yl^2},\label{eq:rhomunu}
\ee (where $g$ is the Landau level degeneracy) whose
time-evolution can be obtained using the Heisenberg equation of
motion (Eq.~(\ref{eq:hem}))\footnote{Note that the density
operators $\rho_{\mu \sigma \nu \sigma'}({\bf k}, \omega)$ are
closely related to the (total) physical density $n({\bf k})$ as $
n({\bf k}) = g e^{-k^2l^2/2}\sum_{\mu\sigma}\rho_{\mu\sigma
,\mu\sigma }({\bf k},0)$, where $g$ is the Landau level
degeneracy.};  the external time-dependent field $\Phi_{\mu \sigma
,\nu \sigma'}({\bf k}, \omega)$ in Eq.~(\ref{eq:pertH1}) must turn
into its complex conjugate when $\mu \sigma \leftrightarrow \nu
\sigma'$ and ${\bf k},\omega \leftrightarrow -{\bf k}, -\omega$ so
that the Hamiltonian remains Hermitian.   Different experimental
probes will couple to different combinations of the density matrix
$\rho_{\mu\sigma ,\nu\sigma' }({\bf k})$.  For example, a surface
acoustic wave experiment might couple to the total charge density
$\sum_{\nu\sigma} \rho_{\nu\sigma,\nu\sigma}$, whereas certain
spin-polarized light scattering experiments might couple to the
spin-flip-density $\sum_{\nu} \rho_{\nu\uparrow, \nu\downarrow}$.

If the perturbing external field $\Phi_{\mu \sigma ,\nu
\sigma'}({\bf k}, \omega)$ is small, one can assume that the
response of the system to it is linear: \be
    \langle \delta \rho_{\mu's,\nu's'}({\bf k}, \omega) \rangle = \chi^{\mbox{\small ret}}_{\mu's,\nu's',\mu\sigma,\nu\sigma'}({\bf k},\omega)\,\, \Phi_{\mu\sigma,\nu\sigma'}({\bf k}, \omega).
\ee The proportionality coefficient between the change in the
density expectation value as a result of the perturbation,
$\langle \delta \rho_{\mu's,\nu's'}({\bf k}, \omega) \rangle$ and
the perturbing field, $\Phi_{\mu \sigma ,\nu \sigma'}({\bf k},
\omega)$, is the density response function that can be obtained in
the TDHF approximation.  The presence of a pole at a particular
frequency and wavevector indicates resonant response (i.e.\ the
presence of a collective mode).

We therefore start the derivation of the collective-mode
dispersion of the $\nu=2$ bilayers by obtaining general
expressions for the density response function. First, in Section
\ref{sec:unres} we review the unrestricted Hartree-Fock
approximation through which the ground states of the $\nu=2$
bilayers were obtained in our previous companion
paper\cite{lopatnikova:03a}.  We then continue in
Section~\ref{sec:drf} to derive the TDHF approximation
\cite{kallin:84,cote:95,wang:02} which is tailored to match the
unrestricted Hartree-Fock of our prior study, and results in a
general matrix equation for the density response function.

In Sections \ref{sec:cmd-00} and \ref{sec:cmd-n0}, collective
modes of the charge-unbalanced $\nu=2$ bilayers in perpendicular
field are obtained (collective modes in charge-balanced $\nu=2$
bilayers were discussed in Ref.~\cite{zheng:97} and
\cite{dassarma:98}).  In Section \ref{sec:cmd-00}, the
collective-mode dispersions (i.e.\ the poles of the density
response function) of the charge-unbalanced $\nu=2$ bilayers with
no interlayer tunneling are obtained in closed form.  The symmetry
properties that simplify the (complicated) general equations for
the density response function are discussed.  In Section
\ref{sec:cmd-n0}, the collective-mode dispersions of the
charge-unbalanced $\nu=2$ bilayers in the presence of a small
amount of interlayer tunneling are obtained numerically and
compared to the collective-mode dispersions of the system without
tunneling (Figs.~\ref{fig:ferro-cmd}--\ref{fig:ss-cmd}).

Section \ref{sec:cmd-t} presents our main result:  the
collective-mode dispersions of the $\nu=2$ bilayer systems in
tilted magnetic field.  A set of collective-mode dispersion curves
calculated as the tilt-angle is increased and the system undergoes
the $C1$--$C2$ transition is exhibited in Fig.~\ref{fig:c1c2-cmd}.
Figure~\ref{fig:cep-cmd} shows the dispersion curves of the
$\nu=2$ bilayers at the critical end-point.  A dramatic softening
of the Goldstone mode at this point is observed.

% LocalWords:  ik xX ik xk yl yl ret zheng dassarma

\section{``Unrestricted'' Hartree-Fock Approximation: An Overview}
\label{sec:unres}

We review here the ``unrestricted" Hatree-Fock first discussed
previously in Refs. \onlinecite{burkov:02} and
\onlinecite{lopatnikova:03a}.   Our system consists of a
disorderless zero temperature bilayer quantum Hall system with
tunneling between the layers and both perpendicular and in-plane
magnetic fields. Three terms of the Hamiltonian --- $H_Z$, the
Zeeman energy, $H_V$ the bias voltage between layers, and  $H_t$
the tunneling --- couple to single electrons and comprise the
non-interacting part of the Hamiltonian \bea
    H_0 &=& H_Z + H_V + H_T \nonumber \\
    &=& - \sum_{X}\mbox{\Large{[}} \Delta_Z S_X^z +  \Delta_V I^z_X + \nonumber \\
        &\mbox{}& + \frac{\Delta_{SAS}}{2} (e^{iQ_{||}X}I^+_X + e^{-iQ_{||}X}I^-_X) \mbox{\Large{]}} \label{eq:ht0}.
\eea where $\vec S$ and $\vec I$ are the spin and isospin
operators \bea
    \vec{S}_X & = & \frac{1}{2} \sum_{\mu\, s s'}c^{\dag}_{\mu s\, X}\vec{\sigma}_{ss'}
    c_{\mu s'\, X}^{\phantom{\dagger}} \\
    \vec{I}_X & = & \frac{1}{2} \sum_{s\, \mu\nu}c^{\dag}_{\mu s\, X}\vec{\tau}_{\mu\nu}
    c_{\nu s\, X}^{\phantom{\dagger}},
\eea where $\vec{\sigma}$ and $\vec{\tau}$ are sets of Pauli
matrices.  Here, the subscript $X$ represents the momentum index
 of the electron state in Landau gauge, and the subscript $s$
 takes on the values $+1$ and
 $-1$ corresponding to spin up and spin down whereas
 $\mu$ and $\nu$ take on the values $+1$ and $-1$ corresponding to different
 layer index (up and down ``isospin").

The Coulomb interactions between the electrons are taken into
account by an additional term \bea
    &&H_I =\frac{1}{2\Omega }\sum_{ \begin{array}{c} \mbox{\small$X_1X_2$} \\
    \mbox{\small$\nu_1,\nu_2$} \\ \mbox{\small$\sigma_1,\sigma_2$} \end{array}}
    \sum_{\bf q}e^{iq_x(X_1-X_2)}e^{-q^2l^2/2}
    V_{\mu_1\mu_2}(q)\times \nonumber\\
    &&\mbox{\hspace*{20pt}}\times c^{\dag}_{\mu_1\sigma_1\, X_1+q_yl^2}
    c^{\dag}_{\mu_2\sigma_2\, X_2}c_{\mu_2\sigma_2\, X_2+q_yl^2}^{\phantom{\dagger}}
    c_{\mu_1\sigma_1\, X_1}^{\phantom{\dagger}}\label{eq:hi}.
\eea where {\it intra-} and {\it inter}layer Coulomb interactions
are \be
    V_{RR}(q) = \frac{2\pi e^2}{\varepsilon q},\ \ V_{RL}(q) = \frac{2\pi e^2}{\varepsilon q}e^{-dq},
\ee respectively, $d$ is the distance between the layers, and
$\Omega$ is the area of the sample.  The total Hamiltonian is
therefore simply \be
    H = H_0 + H_I.\label{eq:H}
\ee

The Coulomb-interacting Hamiltonian in Eq.~(\ref{eq:H}) is not
tractable exactly, and we solve it using the Hartree-Fock
approximation.  In the usual manner, we assume that the many-body
ground state, $|G\rangle$, is a Slater determinant of
single-particle states and perform a functional minimization of
the expectation value $\langle G|H|G\rangle $ with respect to
these single-particle states.  As was described in
Refs.~\onlinecite{burkov:02} and \onlinecite{lopatnikova:03a},
under the assumption of translational invariance in the
$\hat{y}$-direction, the trial ground state $|G\rangle$ can be
written in the form \be
    |G \rangle = \prod_{X} f^{\dag}_{1X}f^{\dag}_{2X}|0\rangle , \label{eq:gt}
\ee where \be
     f_{nX} = \sum_{\mu\sigma }
      (z^n_{\mu\sigma})^* e^{-iQ_{\mu\sigma}X}c_{\mu\sigma X}.\label{eq:f}
\ee A ground state (Eq.~(\ref{eq:gt})) with nonzero
$Q_{\mu\sigma}$ possesses spin-isospin-wave order, discussed at
length in Ref.~\onlinecite{lopatnikova:03a}.

As was mentioned in Ref.~\onlinecite{lopatnikova:03a}, the
proposed ground state  (Eq.~(\ref{eq:f}) and (\ref{eq:gt})) is not
the most general Slater determinant (Hartree-Fock) state. However,
our analysis of the collective modes around the ground states
obtained by the minimization of $\langle G|H|G\rangle$ indicate
the stability of these states against second-order transitions
that cannot be described within the Hilbert space defined by our
ansatz. We note that the possibility of phase transitions into a
soliton-lattice state cannot be ruled out in this work.

To obtain the approximate Hartree-Fock ground state for the
$\nu=2$ bilayer system, we minimize the expectation value of the
Hamiltonian in Eq.~(\ref{eq:H}), $\frac{1}{g}\langle G|H|G
\rangle$, with respect to the variational parameters $z^n_{\mu
\sigma}$, and $Q_{\mu \sigma}$.  As was demonstrated in
Ref.~\onlinecite{lopatnikova:03a}, the resulting set of
minimization conditions can be arranged in the form of a
Schr\"{o}dinger equation: \be
    MZ^n = \epsilon_nZ^n \label{eq:se}
\ee where
$Z^n=(z_{R\uparrow},z_{R\downarrow},z_{L\uparrow},z_{L\downarrow})$,
and $M$ is a $4\times 4$ matrix, which is just the mean-field
single-particle Hartree-Fock Hamiltonian: \bea
    \lefteqn{M_{\nu s';\mu s} = - \Delta_Z \delta_{\mu\nu}\sigma^z_{ss'}- \Delta_V  \delta_{ss'}\tau^z_{\mu\nu} -} \nonumber \\
    && \mbox{} -\Delta_{SAS}\delta_{ss'}\mbox{\Large{[}} \frac{1}{g}\sum_X \cos ((Q_{||}-Q_I)X)\, \tau^x_{\mu\nu} + \nonumber \\
    && \mbox{}+ \frac{1}{g}\sum_X \sin ((Q_{||}-Q_I)X) \, \tau^y_{\mu\nu}\mbox{\Large{]}} +\nonumber \\
    && \mbox{} + 2 H_-\sum_{\mu s }\delta_{\mu\nu}\delta_{ss'}[\sum_{s',m=1,2}|z^m_{\mu s'}|^2-1] - \nonumber \\
    && \mbox{} - F_{\mu\nu} (-[Q_I/2\,(\mu-\nu)+Q_S/2\,(s-s')]\hat{q}_x) \times \nonumber \\
    && \mbox{} \times \sum_{n=1,2}(z^n_{\nu s'})^*z^n_{\mu s} \label{eq:m}.
\eea Here, a further simplification of the problem is made by
making an assumption that \be
    Q_{\mu\sigma} = \frac{\mu}{2}Q_I + \frac{\sigma}{2}Q_S,
\ee where a finite $Q_I$ indicates the presence of an isospin-wave
order, while a finite $Q_S$ reflects the real spin-wave order.
The functions $H_{-} ({\bf q})$ and $F_{\mu\nu} ({\bf q})$ used in
Eq.~(\ref{eq:m}) are defined as \bea
    F_{\mu\nu} ({\bf q}) & = & \int\frac{d^2k}{(2\pi)^2}e^{-k^2l^2/2}V_{\mu\nu}(k) e^{i{\bf q\wedge k} l^2} = \nonumber \\
    & = &\int \frac{dk}{2\pi }e^{-k^2l^2/2} V_{\mu\nu}(k)\, k\, J_0(kql^2) \label{eq:fmunu} \\
    H_{-} ({\bf q})  & = &  \frac{1}{4\pi l^2} (V_{RR}({\bf q})-V_{RL}({\bf q})) = \frac{e^2}{\varepsilon l} \frac{1-e^{-dq}}{2ql},\label{eq:hm}.
\eea These functions arise from the Hatree and exchange parts of
the interaction Hamiltonian (Eq.~(\ref{eq:hi})) treated in the
Hartree-Fock approximation.

The Schr\"{o}dinger equation (Eq.~(\ref{eq:se})) is solved
iteratively \cite{dassarma:98}.  At each iteration, the two
eigenstates corresponding to the lowest eigenvalues are filled
(i.e.\ chosen to be the states 1 and 2).  These lowest-energy
eigenstates $Z^1$ and $Z^2$ are then used to obtain the matrix $M$
for the next iteration.   The procedure is repeated until a
self-consistent solution is achieved.  This solution --- a set of
eigenspinors $Z^n$ sorted according to their eigenvalues ---
defines the lowest energy trial state among the Slater
determinants defined by Eqs.~(\ref{eq:gt}) and (\ref{eq:f})
subject to fixed values of the $Q_{\mu s}$'s. The eigenvalues
$\epsilon_n$ give the binding energy of a particle in the subband
$n$, i.e.\ it is the energy lost when the particle is taken out of
the system. The sum of individual binding energies does not give
the groundstate energy; the ground state energy is calculated from
$\frac{1}{g} \langle G|H|G \rangle $ (see
Ref.~\onlinecite{lopatnikova:03a}). The minimization of the energy
of the ground state over the $Q_{\mu s}$'s is done last
\cite{burkov:02}.  Thus, we find the Hartree-Fock ground state in
two steps: First, for fixed values of $Q_I$ and $Q_S$, we minimize
the expectation value of the ground state energy with respect to
$(z^n_{\mu \sigma})^*$. Then, we minimize the ground state energy
with respect to $Q_I$ and $Q_S$.  Thus, we obtain the phase
diagrams thouroghly discussed in
Ref.~\onlinecite{lopatnikova:03a}, a representative example of
which is given in Fig.~\ref{fig:c1c2single}.

\section{The Density Response Function}
\label{sec:drf}

We start our analysis of the response of the system to external
perturbations by considering the possible excitations of electrons
between the subbands.  The presence of the layer and spin degrees
of freedom of electrons in bilayer systems results in the
splitting of each Landau level into four such subbands.  When the
filling fraction is $\nu=2$, the lowest two subbands are filled,
i.e.\ the Fermi energy lies between the second and the third
subbands.  An elementary transition of a {\it non-interacting}
$\nu=2$ bilayer system occurs when a particle is moved from one of
the filled levels into one of the empty levels, resulting in a
particle-hole pair.  Four such transitions are possible in the
$\nu=2$ bilayers (provided the cyclotron energy is assumed to be
much larger than all the other relevant energy scales):
$1\rightarrow 3$,  $2\rightarrow 4$,  $2\rightarrow 3$,  and
$1\rightarrow 4$. The energy of an {\it unbound} particle-hole
pair is the difference between the energy gained by inserting the
particle into an empty level, $\epsilon_{\beta}$, where $\beta =
3,4$, and the energy lost by removing it from a filled level
$\epsilon_{\alpha}$, where $\alpha=1,2$:  $\Delta\epsilon =
\epsilon_{\beta} - \epsilon_{\alpha}$.  In a real system, the
particles and the holes they leave behind interact.  The
interactions lower the energy of the particle-hole pairs and make
it wavevector dependent.  The energies of the unbound
particle-hole pairs show up as the poles of the HF density
response function, which is, diagrammatically, the bare density
response function dressed with self-energy corrections.  The
self-energy corrections represent the effect of the
renormalization of the single-particle levels by the interactions,
accounted for in the HF approximation.  In order to account for
the particle-hole interactions, the HF density response function
is dressed with vertex corrections.  In the charge-unbalanced
$\nu=2$ bilayers both the Hartree ``bubbles'' and the exchange
``ladders'' contribute to the collective-mode dispersions
\cite{dassarma:98}.

As was discussed in Ref.~\onlinecite{lopatnikova:03a}, we frame
our problem so that the $\nu=2$ bilayer ground state can have a
very simple form, $|G \rangle = \prod_{X}
f^{\dag}_{1X}f^{\dag}_{2X}|0\rangle$, in the basis of the
creation-annihilation operators $f_{nX}$ (Eq.~(\ref{eq:f})).  It
is therefore convenient to define generalized density operators
\be
    \rho_{\alpha \beta } ({\bf k}) =
     \frac{1}{g }\sum_{X} e^{-ik_xX-ik_xk_yl^2/2}f^{\dag }_{\alpha ,X}f_{\beta ,X+k_yl^2}. \label{eq:rhof}
\ee Note that the expectation values of the generalized density
operators are always diagonal in the ground state of the $\nu=2$
bilayer systems, \be
    \langle \rho_{\alpha \beta } ({\bf k})
    \rangle = \delta_{{\bf k},0}\delta_{\alpha\beta}(\delta_{\alpha 1}+\delta_{\alpha 2}).
\ee Since in tilted magnetic fields, in the gauge of our choice,
the operators $f^{\dag}_{\alpha}$ contain the creation operators
$c^{\dag}_{\mu\sigma}$ with different position-dependent phase
factors $e^{iQ_{\mu\sigma}X}$,  the generalized density operators,
$\rho_{\alpha \beta }$, are related to the physical density
operators, $\rho_{\mu\sigma ,\nu\sigma' }$
(Eq.~(\ref{eq:rhomunu})), not only through a linear transformation
but also by a shift of wavevector: \bea
    \rho_{\alpha \beta }({\bf k}) &=&
    \sum_{\mu\nu,\sigma\sigma'}e^{-i(Q_{\mu\sigma} + Q_{\nu\sigma'})k_yl^2/2} z^{\alpha}_{\mu\sigma} (z^{\beta}_{\nu\sigma'})^* \times \nonumber\\
    &&\mbox{}\times \rho_{\mu\sigma ,\nu\sigma' }({\bf k} -
    (Q_{\mu\sigma} - Q_{\nu\sigma'})\hat{x}). \label{eq:rhoab}
\eea

Now, to perturb the system so as to determine its response, we
rewrite the external perturbation Hamiltonian
Eq.~(\ref{eq:pertH1}) in terms of the generalized density
operators as \be
    \delta H = \sum_{\gamma \delta} \int \frac{d\omega}{2 \pi} \int \frac{d^2 k}{2
    \pi}\,\, \,
    \tilde \Phi_{\gamma \delta}({\bf k}, \omega)  e^{-i \omega t}\, \, \rho_{\gamma \delta}^\dagger({\bf
    k}), \label{eq:pertH2}
\ee where \bea
    \lefteqn{\tilde \Phi_{\gamma \delta}({\bf k},\omega) = \sum_{\mu
    \sigma, \nu \sigma'} e^{-i(Q_{\mu\sigma} + Q_{\nu\sigma'})}z^{\alpha}_{\mu\sigma} (z^{\beta}_{\nu\sigma'})^* \times} \nonumber \\
    && \mbox{} \times \Phi_{\mu \sigma, \nu \sigma'}({\bf k} - (Q_{\mu\sigma} - Q_{\nu\sigma'})\hat{x},\omega).
\eea Thus, application of an external potential $\Phi_{\mu \sigma,
\nu \sigma'}$, which couples to the physical density at wavevector
$\bf k$, can generate perturbations $\tilde \Phi$ that couple to
the generalized density at other wavevectors.   For calculational
simplicity, and for simplicity of presenting our results, we will
focus on calculating the response of the system to $\tilde \Phi$
which couples to the generalized density. From this result, one
can simply determine the physical response of the system to an
arbitrary perturbation (in terms of the physical density).
However, these wavevector shifts between the physical density and
the generalized density must be kept in mind as they can be
nontrivial, as we will see below (Sec.~\ref{subsec:cmd-cc} and
Eq.~(\ref{eq:chishift})).

To determine the response of the system to the time-dependent
perturbation in Eq.~(\ref{eq:pertH2}), we use standard linear
response theory (Kubo formula), in which the resulting change in
the expectations of generalized density operator is assumed to be
proportional to the perturbation: \be
    \langle \delta \rho_{\alpha \beta}({\bf k}, \omega) \rangle = \chi^{\mbox{\small ret}}_{\alpha \beta
     \gamma \delta}({\bf k},\omega)\,\, \Phi_{\gamma \delta}({\bf k}, \omega).
\ee The proportionality coefficient is the retarded density
response function \be
    \chi^{\mbox{\small ret}}_{\alpha \beta
     \gamma \delta}({\bf k},\omega)
     = -ig\int_0^{\infty}e^{i\omega t}\langle
     [\rho_{\alpha \beta}({\bf k},t),\rho^{\dag}_{\gamma \delta}({\bf k},0)]
     \rangle. \label{eq:chiret}
\ee We obtain the collective-mode dispersions of the $\nu=2$
bilayers by finding the poles of this response function.

It is convenient to obtain the retarded density response function
(Eq.~(\ref{eq:chiret})) from a corresponding imaginary-time
density response function \be
    \chi_{\alpha \beta \gamma \delta }({\bf k}, \tau ) = -g \langle T \tilde{\rho }_{\alpha \beta }({\bf k},\tau ) \tilde{\rho }^{\dag }_{\gamma \delta}({\bf k},0)\rangle, \label{eq:chiem}
\ee where $\tilde{\rho}_{\alpha \beta } = \rho_{\alpha \beta
}-\langle \rho_{\alpha \beta } \rangle $.  The imaginary-time
density response function can be Matsubara-transformed to get
$\chi_{\alpha \beta \gamma \delta }({\bf k}, i\Omega )$ (where
$i\Omega_n $ are bosonic frequencies) that, in turn, can be
transformed into the retarded density-response function
$\chi^{\mbox{\small ret}}_{\alpha \beta \gamma \delta }({\bf k},
\omega)$ by a Wick rotation $i\Omega \to \omega +i\delta $.

Following C\^{o}t\'{e} and MacDonald (CM) \cite{cote:91}, we
proceed by calculating the Hartree-Fock density response function
$\chi^0_{\alpha \beta \gamma \delta }({\bf k}, i\Omega )$ from its
equation of motion: \bea
    \lefteqn{-\frac{1}{g}\frac{d}{d\tau}\chi_{\alpha \beta \gamma \delta }^0({\bf k}, \tau ) = } \\
    && = \delta (\tau ) \langle [\rho_{\alpha \beta }({\bf k},0), \rho^{\dag }_{\gamma \delta}({\bf k},0)] \rangle + \langle T \frac{\partial }{\partial \tau} \rho_{\alpha \beta }({\bf k},\tau) \rho^{\dag }_{\gamma \delta}({\bf k},0)\rangle \nonumber.
\eea The commutation relations of the generalized density
operators $\rho_{\alpha \beta }({\bf k},0)$ are \be
    g[\rho_{\alpha \beta }({\bf k}), \rho^{\dag }_{\gamma \delta}({\bf k})] = \delta_{\beta \delta}\rho_{\alpha\gamma}(0)- \delta_{\alpha\gamma}\rho_{\beta\delta}(0).
\ee The time-evolution of the density operator is determined,
within the Hartree-Fock approximation, from the mean-field
Hartree-Fock Hamiltonian, ${\cal H}^{HF}$: \bea
     \lefteqn{\frac{\partial }{\partial \tau} \rho_{\alpha \beta }({\bf k},\tau) = [{\cal H}^{HF}, \rho_{\alpha \beta }({\bf k},\tau)] = }\label{eq:hem}\\
    && = e^{i({\cal H}^{HF}-\mu N)\tau }  [{\cal H}^{HF}, \rho_{\alpha \beta }({\bf k},0)] e^{-i({\cal H}^{HF}-\mu N)\tau },\nonumber
\eea where the mean-field Hamiltonian is diagonal in the
$f_{\alpha}$-basis, and can simply be written as (see
Sec.~\onlinecite{lopatnikova:03a}) \be
    {\cal H}^{HF} = \sum_{\alpha }\epsilon_{\alpha}f^{\dag}_{\alpha}f_{\alpha}.
\ee Using the Hartree-Fock equation of motion for the density
operators, and Matsubara-transforming the equation of motion for
the density response function, we get \bea
    \lefteqn{i\Omega_n \chi^0_{\alpha \beta \gamma \delta }({\bf k},i\Omega_n ) = }\label{eq:chi0}\\
    && = \delta_{\beta \delta }\langle \rho_{\alpha \gamma }(0) \rangle - \delta_{\alpha \gamma }\langle \rho_{\delta \beta }(0) \rangle + (\epsilon_{\beta}-\epsilon_{\alpha})\chi^0_{\alpha \beta \gamma \delta }({\bf k}, i\Omega_n).\nonumber
\eea The single-particle density response function is therefore
\be
    \chi^0_{\alpha \beta \gamma \delta }({\bf k},i\Omega_n ) = \frac{\delta_{\alpha\gamma}\delta_{\beta\delta}\, \mbox{sign} (\epsilon_{\beta}-\epsilon_{\alpha})}{i\Omega_n-\epsilon_{\beta}+\epsilon_{\alpha}}.
\ee Its poles are, as expected, at the single-electron excitation
energies.

To take into account the interactions between the single-particle
excitations, following Refs.~\onlinecite{cote:91} and
\onlinecite{kallin:84}, we introduce the vertex corrections.  The
vertex corrections in TDHF are Hartree ``bubbles'' and exchange
``ladders'', which have to be related to the Hartree-Fock
self-energies through Ward identities \cite{dassarma:98}.  Using
the interaction constants \bea
    H_{\alpha \beta \gamma \delta}( {\bf k}) && = \frac{1}{2\pi l^2} \sum_{i_1,i_2,\sigma_1,\sigma_2} z^{\alpha}_{i_1\sigma_1} (z^{\beta}_{i_1\sigma_1})^* z^{\gamma}_{i_2\sigma_2} (z^{\delta}_{i_2\sigma_2})^* \times \nonumber \\
    && \mbox{} \times V_{i_1i_2}(k) e^{-k^2l^2/2} e^{-i k_y (Q_{i_1\sigma_1}-Q_{i_2\sigma_2})l^2} \label{eq:hmat} \\
    F_{\alpha \beta \gamma \delta}( {\bf k}) && = \sum_{i_1,i_2,\sigma_1,\sigma_2} z^{\alpha}_{i_1\sigma_1} (z^{\beta}_{i_1\sigma_1})^* z^{\gamma}_{i_2\sigma_2} (z^{\delta}_{i_2\sigma_2})^* \times \nonumber \\
    && \mbox{} \times F_{i_1i_2}([k_x-(Q_{i_1\sigma_1}-Q_{i_2\sigma_2})]\hat{x}+k_y\hat{y})\label{eq:fmat}
\eea we dress the single-particle density response function to
include the interactions between the single-particle excitations:
\bea
     \lefteqn{\tilde{\chi}_{\alpha \beta \gamma \delta }({\bf k},i\Omega_n ) =  \chi^0_{\alpha \beta \gamma \delta }({\bf k},i\Omega_n ) -}\nonumber \\
    && \mbox{} - \chi^0_{\alpha \beta a b}({\bf k},i\Omega_n ) F_{cabd}( {\bf k})  \tilde{\chi}_{c d \gamma \delta }({\bf k},i\Omega_n ) \\
    \lefteqn{\chi_{\alpha \beta \gamma \delta }({\bf k},i\Omega_n ) = \tilde{\chi}_{\alpha \beta \gamma \delta }({\bf k},i\Omega_n ) +}\nonumber \\
    && \mbox{} + \tilde{\chi}_{\alpha \beta ab }({\bf k},i\Omega_n ) H_{bacd}( {\bf k})\chi_{cd\gamma \delta }({\bf k},i\Omega_n ); \label{eq:chi}
\eea summation over repeated indices is implied, and the exchange
interaction functions $F_{i_1i_2}({\bf k})$ are defined in
Eq.~(\ref{eq:fmunu}).

To solve for the dressed density response function, $\chi_{\alpha
\beta \gamma \delta }({\bf k},i\Omega_n )$, it is convenient to
cast Eqs.~(\ref{eq:chi0})--(\ref{eq:chi}) into matrix form.  With
the definitions \bea
    X_{4\alpha +\beta -5,4\gamma +\delta -5} & = & \chi_{\alpha \beta \gamma \delta }({\bf k},i\Omega_n ), \\
    R_{4\alpha +\beta -5,4\gamma +\delta -5} & = & \delta_{\beta \delta }\langle \rho_{\alpha \gamma }(0) \rangle - \delta_{\alpha \gamma }\langle \rho_{\delta \beta }(0) \rangle \nonumber \\
    & = & \delta_{\alpha \gamma } \delta_{\beta \delta } \sum_{n=1,2}(\delta_{\alpha ,n}-\delta_{\beta ,n}), \\
    M_{4\alpha +\beta -5,4\gamma +\delta -5} & = & \delta_{\alpha \gamma } \delta_{\beta \delta}(\epsilon_{\alpha } - \epsilon_{\beta }), \\
    H_{4\alpha +\beta -5,4\gamma +\delta -5} & = & H_{\beta \alpha \gamma \delta} ({\bf k}), \\
    F_{4\alpha +\beta -5,4\gamma +\delta -5} & = & F_{\gamma \alpha \beta \delta} ({\bf k}),
\eea we have \bea
    i\Omega_n X_0 & = & R - M X_0, \\
    \tilde{X} & = & X_0 - X_0 F \tilde{X}, \\
    X & = & \tilde{X} - \tilde{X} H X.
\eea The density response function $\chi_{\alpha \beta \gamma
\delta }({\bf k},i\Omega_n)$ is represented by the $16\times 16$
matrix X: \be
    X = [i\Omega_n - M - R(H-F)]^{-1} R.\label{eq:chimat}
\ee The poles of the density response function are the solutions
to the secular equation \be
    \det [i\Omega_n - M -R(H-F)] = 0.\label{eq:findpoles}
\ee

The $16\times 16$ matrix equations can be reduced to $8\times 8$
by eliminating the forbidden single-particle excitations, such as
$2 \rightarrow 2$, or $3 \rightarrow 4$.  Even though the
effective ``energies'' of these transitions are solutions to the
secular equation (Eq.~(\ref{eq:findpoles})), it is easy to show
that the weights of these modes are always 0 and they do not show
up in the density-response function matrix.  The remaining $8
\times 8$ matrix equation includes the interactions between the
four transitions that create the particle-hole pairs, and four
corresponding transitions that recombine the particles and the
holes.

% LocalWords:  diagrammatically renormalization DSSZ TDHF ret ig ik xX xk yl iQ
% LocalWords:  CHIS Matsubara KH cabd ab bacd cd subband SAS Shr cccc subbands
% LocalWords:  decoupled gapless RL ql superpositions FERRO pre destabilized nX
% LocalWords:  decouples analyticity energetics Larmor calculational kallin

% LocalWords:  zheng dassarma leaves superfluid superfluid superfluid
% LocalWords:  superfluid superfluids anisotropies
\section{Collective-mode dispersions of charge-unbalanced $\nu=2$ bilayers, $\Delta_{SAS} = 0$}
\label{sec:cmd-00}

In the most general case, the solution to Eq.~(\ref{eq:findpoles})
has to be found numerically.  The situation simplifies
considerably when either the bias voltage or the tunneling is
zero.  The former case has been studied in Refs.~\cite{zheng:97}
and \cite{dassarma:98}, who obtained the spin-density wave
branches of the collective modes in perpendicular magnetic field
(their analysis can be easily extended to the case of tilted
fields).  The latter case is considered in this section:  Using
the symmetry properties of the $\nu=2$ bilayers in the absence of
tunneling, we show that different inter-subband single-particle
excitations are independent of each other in $\nu=2$ bilayer
systems in the absence of tunneling and the vertex corrections
simply result in additional ($q$-dependent) renormalization of the
excitations.

In this section, we present the analytical calculation of the
collective-mode dispersions in $\nu=2$ bilayer systems in the
absence of tunneling in perpendicular field.  We explain the main
features of the dispersion curves and, in the second part of this
section, discuss the evolution of these features as the interlayer
tunneling is turned on.  The $\nu=2$ bilayers in titled magnetic
fields are considered in the next section.

\subsection{Parametrized $\Delta_{SAS} = 0$ ground state}
\label{subsec:paramgs}

We start the calculation  of the collective-mode dispersions by
finding the ground state of the system.  As was discussed in
Sec.~\ref{sec:unres}, the ground state of the $\nu=2$ bilayers is
obtained within the Hartree-Fock approximation by solving the
Schr\"{o}dinger-like equation (Eq.~(\ref{eq:se})).  In the absence
of tunneling, the mean-field solutions $Z^n$ can be parametrized
by two parameters, so that a transformation matrix, $S$, that can
be constructed of the four eigenspinors, $Z^n$, has the form: \be
    S = (Z^1,Z^2,Z^3,Z^4)
    = \left( \begin{array}{cccc}
                1 & 0 & 0 & 0 \\
                0 & \sin\theta & e^{i\phi}\cos\theta & 0 \\
                0 & e^{-i\phi} \cos\theta & -\sin\theta & 0 \\
                0 & 0 & 0 & 1
                \end{array} \right),\label{eq:smat}
\ee where $Z^n$ are defined after Eq.~(\ref{eq:se}).   The two
subbands with the lowest binding energies are filled.  By
construction, for positive bias voltage and Zeeman coupling, the
lowest are the bands 1 and 2, so that the general form of the
ground state is $\prod_X c^{\dag}_{R\uparrow X} (\sin\theta \,
c^{\dag}_{R\downarrow X} +e^{i\phi}\cos\theta\,
z_{L\uparrow}c^{\dag}_{L\uparrow X})|0\rangle $.  When $\cos\theta
= 1$, the ground state is the ferromagnetic state;  when
$\cos\theta = 0$, it is the spin-singlet state; the intermediate
values of $\cos\theta$ indicate that the system is in the
 many-body so-called $I$-state (See Ref.~onlinecite{lopatnikova:03a}).
 It is easy to see that the
Hamiltonian is invariant with respect to change of the phase
$\phi$ --- this is the $U(1)$ symmetry that results in the
formation of a Goldstone mode in the $I$-phase.  To simplify our
calculations, we choose $\phi = 0$, so that the matrix $S$ is now
real and $S=S^{-1}$.

When the parameter $\theta$ is such that the ground state energy
is minimized, the $S$ matrix diagonalizes the mean-field
Hamiltonian matrix $M$ (recall that the Schr\"{o}dinger-like
equation (Eq.~(\ref{eq:se})) is the result of a formal
minimization of the Hartree-Fock ground state energy with respect
to the parameters $(z^n_{\mu\sigma})^*$). We can therefore find
$\theta$ by forcing the matrix $\Lambda = S^{-1}MS$ to be diagonal
--- i.e.\@ equating the off-diagonal terms of the matrix $\Lambda$
to 0.  The resulting minimization condition can be written as \be
    K_0(\theta) \sin 2\theta = 0.\label{eq:sc}
\ee where the function $K_0(\theta)$ is defined as \be
    K_0(\theta) = -\frac{\Delta_V-\Delta_Z-2H_-}{2}
    + (H_- - F_-)\cos 2\theta .\label{eq:kappa}
\ee Equation (\ref{eq:sc}) is satisfied automatically in the
ferromagnetic and spin-singlet phases, where $\cos\theta =0$ and
$1$, respectively, so that $\sin 2\theta = 0$.  In the $I$-phase,
the equation is solved by \be
    \cos^2\theta=\frac{1}{2}-\frac{\Delta_V-\Delta_Z-2H_-}{4(H_--F_-)},\label{eq:sc2}
\ee where \bea
    F_{\pm} & = & \frac{1}{2}\int\frac{d^2q}{(2\pi)^2}e^{-q^2l^2/2}(V_{RR}(q)\pm V_{RL}(q)) = \nonumber \\
    && = \frac{e^2}{\varepsilon l} \frac{1}{2}\sqrt{\frac{\pi }{2}}\left[ 1 \pm e^{d^2/2l^2}\mbox{Erfc}\left(\frac{d}{\sqrt{2}l}\right) \right],\label{eq:fpm}
\eea and $H_-$ is defined in Eq.~(\ref{eq:hm}).
Equation~(\ref{eq:sc2}) has a solution when its right-hand side
takes a value between 0 and 1.  This region is the region of
stability of the $I$-phase. (Note that, in the absence of
anisotropy between the {\it inter-} and {\it intra}layer
interactions, Eq.~(\ref{eq:sc2}) would have no solution, since
$H_- =0$ and $F_-=0$ in this case.)  If the right-hand side of
Eq.~(\ref{eq:sc2}) is negative, the ground state energy is
minimized by $\cos^2\theta =0$ and the system is in the
ferromagnetic phase.  If the right-hand side of Eq.~(\ref{eq:sc2})
is greater than 1, then  $\cos^2\theta =1$ and the system is in
the spin-singlet phase.  Intermediate values of $\cos^2\theta $
give the $I$-state.  Note that the function $K_0(\theta)=0$
throughout the $I$-phase and takes on finite values in the
ferromagnetic and spin-singlet phases.

The resulting binding energies are the eigenvalues of the matrix
$M$, and can be read off the diagonal of the matrix $\Lambda =
S^{-1}MS$. \bea
    \epsilon_1 &=& -\Delta_Z+F_-\cos 2\theta +K_0(\theta) -(F_++F_-)\label{eq:e1}\\
    \epsilon_2 &=& F_-+K_0(\theta)\cos 2\theta -(F_++F_-)\label{eq:e2}\\
    \epsilon_3 &=& -F_--K_0(\theta)\cos 2\theta \label{eq:e3}\\
    \epsilon_4 &=& \Delta_Z -F_-\cos 2\theta-K_0(\theta)\label{eq:e4}.
\eea Note that, in the many-body region, where $K_0(\theta) = 0$,
the smallest single-particle gap depends only on the interaction
constants, $\epsilon_3-\epsilon_2 = F_+-F_-$, and is constant
throughout the region.  This is again a manifestation of the
many-body nature of the $I$-phase.

\subsection{Symmetry properties of the $\Delta_{SAS} = 0$ ground states}
\label{subsec:symm}

In Ref.~\onlinecite{lopatnikova:03a}, we pointed out that the
ground states realized in $\nu=2$ bilayer systems in the absence
of tunneling and with positive Zeeman field and bias voltage are
eigenstates of the operator $I^z+S^z$ with the eigenvalue $g$
(where $g$ is the Landau level degeneracy, i.e.\ $+1$ per flux
quantum).   The operator $I^z+S^z$ commutes with the bilayer
Hamiltonian and therefore provides a good quantum number to
classify the eigenstates of the Hamiltonian.  Thus, the ground
state belongs to the class of states with the $I^z+S^z$-quantum
number equal to $g$.  So does the lowest-energy excited state,
which is the result of the $2 \rightarrow 3$ transition;  that is
to say, the transition $2 \rightarrow 3$ does not change the
$I^z+S^z$-quantum number of the system, $\delta(I^z+S^z) = 0$. The
excitations $1\rightarrow 3$ and $2\rightarrow 4$, on the other
hand, lower the $I^z+S^z$-quantum number by one, i.e.\
$\delta(I^z+S^z) = -1$, and the highest-energy excitation
$1\rightarrow 4$ lowers the $I^z+S^z$-quantum number by
$\delta(I^z+S^z) = -2$.

Modes characterized by the same $\delta(I^z+S^z)$ can be mixed by
the Coulomb interactions, unless they can be classified further by
other quantum numbers.  For the excitations  $1\rightarrow 3$ and
$2\rightarrow 4$, simply the operator $n_{R\uparrow} =
c^{\dag}_{R\uparrow}c_{R\uparrow}$ provides a good quantum number
(it commutes with the zero-tunneling Hamiltonian): the excitation
$2\rightarrow 4$ possesses the same eigenvalue $g$ as the ground
state, while the excitation $1\rightarrow 3$ changes it by $\delta
n_{R\uparrow} = -1$.  Therefore, all the four single-particle
inter-subband modes are decoupled from each other as states with
different conserved quantum numbers.  Only the particle-hole
interactions within the same mode therefore appear in the
calculation of the collective modes, and matrix of the density
response function thus separates into four $2\times 2$ matrices.

\subsection{$\Delta_{SAS}=0$ collective-mode dispersions --- general}
\label{subsec:cmd-00g}

Reduced to include only one excitation mode $\alpha \rightarrow
\beta$, and its counterpart $\beta \rightarrow \alpha$, the
$2\times 2$ density-response-function matrix obeys a matrix
equation of the same general form as the full equation
(Eq.~(\ref{eq:chimat})).  The matrices comprising the reduced
Eq.~(\ref{eq:chimat}) are \bea
    &&M = \left( \begin{array}{cc}
                \epsilon_{\beta} -\epsilon_{\alpha} & 0\\
                0 & - (\epsilon_{\beta}-\epsilon_{\alpha})
                \end{array} \right) \\
    &&R = \left( \begin{array}{cc}
                1 & 0\\
                0 & -1
                \end{array} \right).
\eea The Hartree part of the vertex-correction matrix is \be
    H = \left( \begin{array}{cc}
                H_{\beta\alpha\alpha\beta} & H_{\beta\alpha\beta\alpha}\\
                H_{\alpha\beta\alpha\beta} & H_{\alpha\beta\beta\alpha}
                \end{array} \right)
    = \left( \begin{array}{cc}
                H_0 & H_0\\
                H_0 & H_0
                \end{array} \right),
\ee where $H_0 = H_{\beta\alpha\alpha\beta}$ and we used the
symmetries of $H_{\alpha\beta\gamma\delta}$ in Eq.~(\ref{eq:hmat})
with respect to the exchange of indices (given that, without loss
of generality, the coefficients $z^{\alpha}_{i\sigma}$ can be
assumed to be real in the absence of tunneling).  We use the same
symmetries (Eq.~(\ref{eq:fmat})) to obtain the general form of the
exchange contribution to the vertex corrections: \be
    F = \left( \begin{array}{cc}
                F_{\alpha\alpha\beta\beta} & F_{\beta\alpha\beta\alpha}\\
                F_{\alpha\beta\alpha\beta} & F_{\beta\beta\alpha\alpha}
                \end{array} \right)
    = \left( \begin{array}{cc}
                F_0 & F_1\\
                F_1 & F_0
                \end{array} \right),
\ee where $F_0 = F_{\alpha\alpha\beta\beta}$ and $F_1 =
F_{\alpha\beta\alpha\beta}$.  The solution to
Eq.~(\ref{eq:findpoles}) gives the dispersion curve of the
collective excitation $\alpha \rightarrow \beta$: \bea
    \lefteqn{\omega_{\alpha\beta} = \sqrt{F_0-F_1-(\epsilon_{\beta}-\epsilon_{\alpha})}\times } \label{eq:om}\\
    &&\mbox{} \times \sqrt{-2H_0+F_0+F_1-(\epsilon_{\beta}-\epsilon_{\alpha})}.\nonumber
\eea where  the $\epsilon_\alpha$  are given in
Eqs.~(\ref{eq:e1})-(\ref{eq:e4}).

The resulting collective-mode dispersions are given in
Figs.~\ref{fig:ferro-cmd}--\ref{fig:ss-cmd}.

\subsection{Goldstone mode}
\label{subsec:gm}

We start by considering the lowest-energy mode, which softens when
the system enters the $I$-phase.  Using the parametrization of the
coefficients $z^n_{i\sigma}$ given by Eq.~(\ref{eq:smat}), we get
\bea
    &&H_0 = H_{3223} = H_-\sin^2 2\theta \\
    &&F_0 = F_{2233} = F_+-F_-\cos^22\theta \\
    &&F_1 = F_{2323} = F_-\sin^2 2\theta .
\eea The resulting collective mode dispersion is \be
    \omega_{23} =\sqrt{\Theta (\theta ,q)\,(\Theta (\theta ,q)+\Gamma (\theta ,q))} \label{eq:o23}\\
\ee where \bea
    &&\Theta (\theta ,q) = 2K_0(\theta )\cos 2\theta+\tilde{F}_-(q)-\tilde{F}_+(q) \nonumber \\
    &&\Gamma (\theta ,q) = 2(H_-(q)-F_-(q))\sin^2 2\theta \nonumber,
\eea and $\tilde{F}_{\alpha}(q) = F_{\alpha}(q) -  F_{\alpha}(0)$.
($F_{\pm}({\bf q}) = \frac{1}{2}(F_{RR}({\bf q})\pm F_{RL}({\bf
q}))$, where $F_{\mu\nu}({\bf q})$ is defined in
Eq.~(\ref{eq:fmunu}); $H_-({\bf q})$ is defined in
Eq.~(\ref{eq:hm}))

The functions $\tilde{F}_{\alpha}(q)$ are proportional to $q^2$ at
small values of $q$, while $H_-$ approaches a finite constant.
When the system is in the $I$-phase, where $K_0(\theta ) =0$ and
$\sin^2 2\theta \neq 0$, the dispersion curve becomes gapless and
$\omega_{23} \propto |q|$ at small $q$.  Indeed, this is the
linearly dispersing Goldstone mode that appears in the $I$-phase
as a result of the spontaneously broken $U(1)$ symmetry of the
ground state.  The Goldstone mode disperses linearly since the
generator of symmetry does not commute with the Hamiltonian.  The
velocity of the Goldstone mode is \be
    v_G = l\sin 2\theta \sqrt{2(H_--F_-)}\sqrt{\left[ \frac{d}{2l}\frac{e^2}{\varepsilon l}-(1+\frac{d^2}{l^2})F_{RL}\right]},
\ee where $F_{RL} = F_+ -F_-$ is calculated in Eq.~(\ref{eq:fpm}).
The Goldstone-mode velocity is proportional to $\sin 2\theta$ ---
it is zero at the phase boundaries and the greatest near the
middle of the $I$-phase.

In the ferromagnetic and spin-singlet phases, the dispersion curve
of the mode is gapped and analytical around $q=0$.  In the
ferromagnetic state the dispersion is \be
    \omega^F_{23} = \Delta_Z+2F_--\Delta_V +(\tilde{F}_-(q)-\tilde{F}_+(q)),
\ee consistently with the fact that the ferromagnetic state is
stabilized by the magnetic field and by the anisotropy of the
Coulomb interaction in bilayer systems.  As the bias voltage is
increased, the gap is reduced, until it becomes zero when
$\Delta_V = \Delta_Z+2F_-$.  The transition to the $I$-state
occurs at this point, consistently with Eq.~(\ref{eq:sc2}). The
dispersion of the $2\rightarrow 3$ mode in the spin-singlet phase
is \be
    \omega^S_{23} = \Delta_V - \Delta_Z - 4H_- + 2F_- + (\tilde{F}_-(q)-\tilde{F}_+(q)).\label{eq:om23s}
\ee In the absence of tunneling, the system is driven into the
spin-singlet phase by the external bias voltage and against the
renormalized interlayer charging energy $2(2H_- - F_-)$.  Again,
consistently with Eq.~(\ref{eq:sc2}), the system undergoes a
mode-softening phase transition between the spin-singlet phase and
the $I$-phase when $\Delta_V = \Delta_Z +2(2H_- - F_-)$.

The last interesting feature of the $2\rightarrow 3$ mode we
consider is the roton minimum that this mode develops in the
$I$-phase (see Fig.~\ref{fig:canted-cmd}).  The roton minimum
appears deep in the $I$-phase and disappears close to the
boundaries with the ferromagnetic and spin-singlet phases.  It
occurs at $ql \approx 1$, a wavevector characteristic of
interaction effects \cite{cote:95}. In the present case, the roton
minimum indicates a tendency toward formation of an interlayer
spin-density wave.  (Formally, it is the non-trivial wavevector
dependence of $H_-(q)-F_-(q)$ in Eq.~(\ref{eq:o23}) that causes
the roton minimum to appear.)

\subsection{Spin-wave modes}
\label{subsec:sw}

The dispersions of the modes $1\rightarrow 3$ and $2\rightarrow 4$
have the form given by Eq.~(\ref{eq:om}).  It is clear from the
parametrization of the $z^n_{\mu\sigma}$ (Eq.~(\ref{eq:smat}))
that $H_0=F_1=0$ for the excitations $1\rightarrow 3$ and
$2\rightarrow 4$.  The $2\times 2$ interaction matrices $H$ and
$F$ are therefore diagonal, and the dispersion curves have the
simple form \be
    \omega = (\epsilon_{\beta}-\epsilon_{\alpha})-F_0, \label{eq:od}
\ee Unlike the dispersion of the Goldstone mode $2\rightarrow 3$,
the dispersions of the $1\rightarrow 3$ and $2\rightarrow 4$ modes
are analytical in all the phases of the $\nu=2$  bilayers in the
absence of tunneling. The only relevant interaction constant,
$F_0$, is the same for both modes, $1 \rightarrow 3$ and
$2\rightarrow 4$, \be
    F_0 = F_{1133}=F_{2244} = F_+(q)-F_-(q)\cos 2\theta ,
\ee and so are the binding-energy differences,
$\epsilon_3-\epsilon_1 = \epsilon_4-\epsilon_2$.  The
collective-mode dispersions are therefore degenerate, and \bea
    \lefteqn{\omega_{13} = \omega_{24} = }\label{eq:om-b}\\
    && = \Delta_Z-\tilde{F}_+(q)-\tilde{F}_-(q)\cos 2\theta -2K_0(\theta)\sin^2\theta \nonumber .
\eea The dispersion is always gapped; the gap is equal to the
Zeeman splitting in the ferromagnetic and the $I$-phases.  The
system possesses a finite magnetization in these phases, so that
the modes $1\rightarrow 3$ and $2\rightarrow 4$ correspond to
spin-wave modes.  In the spin-singlet phase the magnetization is
zero, and the gap of $1\rightarrow 3$ and $2\rightarrow 4$ modes
departs from the Zeeman splitting linearly with $\Delta_V$: \bea
    \lefteqn{\omega^S_{13} = \omega^S_{24} = }\\
    && = \Delta_V -4H_- + 2F_- + (\tilde{F}_-(q)-\tilde{F}_+(q))\nonumber .
\eea That the gap of $\omega^S_{13}$ is equal to the Zeeman
splitting at the boundary of the spin-singlet and the $I$-phases
is more clear if one compares this dispersion curve to
$\omega^S_{23}$, given in Eq.~(\ref{eq:om23s}).  The difference
between $\omega^S_{13}$ and $\omega^S_{23}$ equals to the Zeeman
splitting at any $\Delta_V$ and $q$.  This is because the excited
state produced by the excitation $2 \rightarrow 3$ is a
spin-triplet state with $S^z =+1$.  The excited states that result
from $1\rightarrow 3$ and $2\rightarrow 4$ are superpositions of a
spin-singlet state and a spin-triplet state with $S^z =0$.  The
excitation $1\rightarrow 4$, which we consider below, in the
spin-singlet state results in a spin-triplet excited state with
$S^z =-1$.

The degeneracy of the modes $1\rightarrow 3$ and $2\rightarrow 4$
is a consequence of the up-down, left-right symmetry of the
Hamiltonian that reverses the sequence of subbands:  the
transformation that exchanges the levels 1 and 4, and 2 and 3 ---
$\Delta_Z \rightarrow -\Delta_Z ,\ \Delta_V \rightarrow
-\Delta_V,\ \Delta_{SAS} \rightarrow -\Delta_{SAS},$ and
$c_{\mu\sigma} \rightarrow e^{i\mu\pi/2}c_{-\mu\, -\sigma}$ leaves
the Hamiltonian invariant.  This symmetry maps the transition
$1\rightarrow 3$ to $4\rightarrow 2$.

\subsection{Highest-energy mode}
\label{subsec:hem}

For the highest-energy mode, $1 \rightarrow 4$, the interaction
matrices also turn out to be diagonal, $H_0 = F_1 = 0$ in
Eqs.~(\ref{eq:hmat}) and (\ref{eq:fmat}).  The collective-mode
dispersion therefore has the same form as that for the
$1\rightarrow 3$ and $2\rightarrow 4$ modes (Eq.~(\ref{eq:om-b})).
The remaining exchange interaction constant, $F_0$, is \be
    F_0=F_{1144} = F_+(q)-F_-(q),
\ee and the dispersion relation, therefore, is \bea
    \lefteqn{\omega_{14} = \Delta_Z+\Delta_V-2(H_--F_-)-} \label{eq:om14}\\
    && \mbox{} - 2H_-\cos 2\theta -\tilde{F}_+(q)+\tilde{F}_-(q)\nonumber .
\eea When $\Delta_V = 0$, the mode $1\rightarrow 4$ is degenerate
with the $2\rightarrow 3$ mode (Eq.~(\ref{eq:om23s})).  The modes
are split when a bias voltage is applied to the system, and the
splitting grows as $2\Delta_V$ until the system enters the
$I$-phase (at the point where the $2\rightarrow 3$ dispersion
becomes gapless).  The gap of the $1\rightarrow 4$ mode starts
decreasing as the system is brought deeper into the $I$-phase.  At
the boundary between the $I$-phase and the spin-singlet phase, the
gap is at its lowest value of $2\Delta_Z$.  In the spin-singlet
phase, the dispersion is \be
    \omega^S_{14} = \Delta_V + \Delta_Z - 4H_- + 2F_- + (\tilde{F}_-(q)-\tilde{F}_+(q)),\label{eq:om14s}
\ee and continues linearly with $\Delta_V$, as it did in the
ferromagnetic phase.  The dispersion $\omega^S_{14}$ is simply
related $\omega^S_{23}$ and $\omega^S_{13}$, as the excitation
resulting in the third ($S^z = -1$) of the triplet excited states.
\begin{figure}[t]
    \centering
    \scalebox{0.65}{\includegraphics{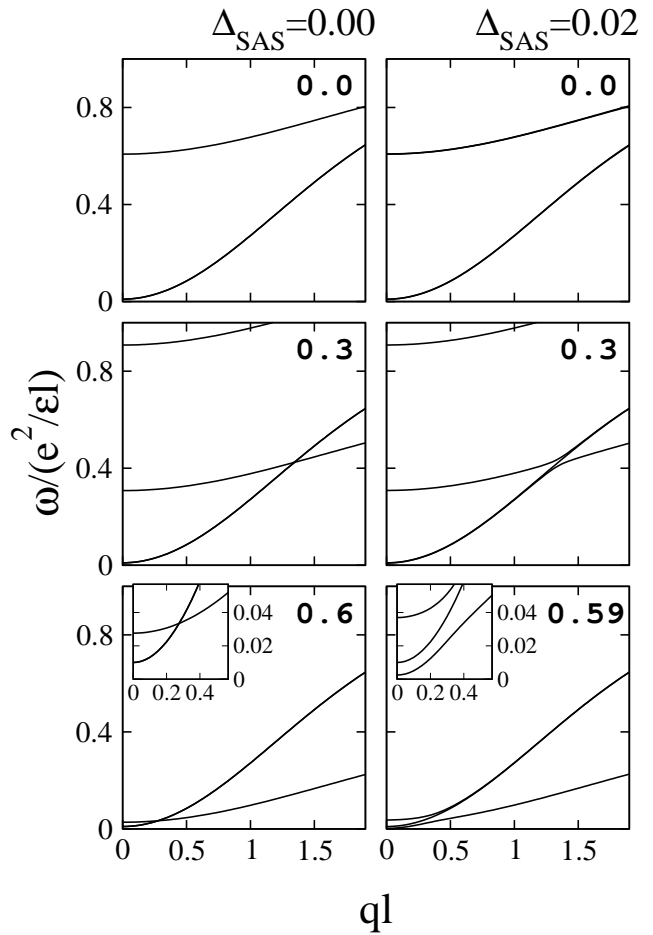}}
    \caption[The collective-mode dispersions of the charge-unbalanced
    $\nu=2$ bilayers in perpendicular field --
    {\it ferromagnetic phase}.]{\label{fig:ferro-cmd}
    The collective-mode dispersions of the charge-unbalanced
    $\nu=2$ bilayers in perpendicular field --
    {\it ferromagnetic phase}. The dispersions
    in the left column are given for a system
    with $\Delta^0_{SAS} = 0.0(e^2/\varepsilon l)$;
    in the right column -- for a system with
    $\Delta^0_{SAS} = 0.02(e^2/\varepsilon l)$;
    the Zeeman energy in all panels is $\Delta_Z = 0.01
    (e^2/\varepsilon l)$; the bias voltage is given in the
    upper right corner of each panel in
    units of $(e^2/\varepsilon l)$.  Some collective-mode dispersions
    are degenerate.  The low-energy, low-wavevector region is shown in the insets.}
\end{figure}

% LocalWords:  diagrammatically renormalization DSSZ TDHF ret ig ik xX xk yl iQ
% LocalWords:  CHIS Matsubara KH cabd ab bacd cd subband SAS Shr cccc subbands
% LocalWords:  decoupled gapless RL ql superpositions FERRO pre destabilized nX
% LocalWords:  decouples analyticity energetics Larmor calculational kallin

% LocalWords:  zheng dassarma leaves superfluid superfluid superfluid
% LocalWords:  superfluid superfluids anisotropies

\section{Collective-mode dispersions of charge-unbalanced $\nu=2$ bilayers, $\Delta \neq 0$, $Q_{||}=0$}
\label{sec:cmd-n0}

\begin{figure}[t!]
    \centering
    \scalebox{0.65}{\includegraphics{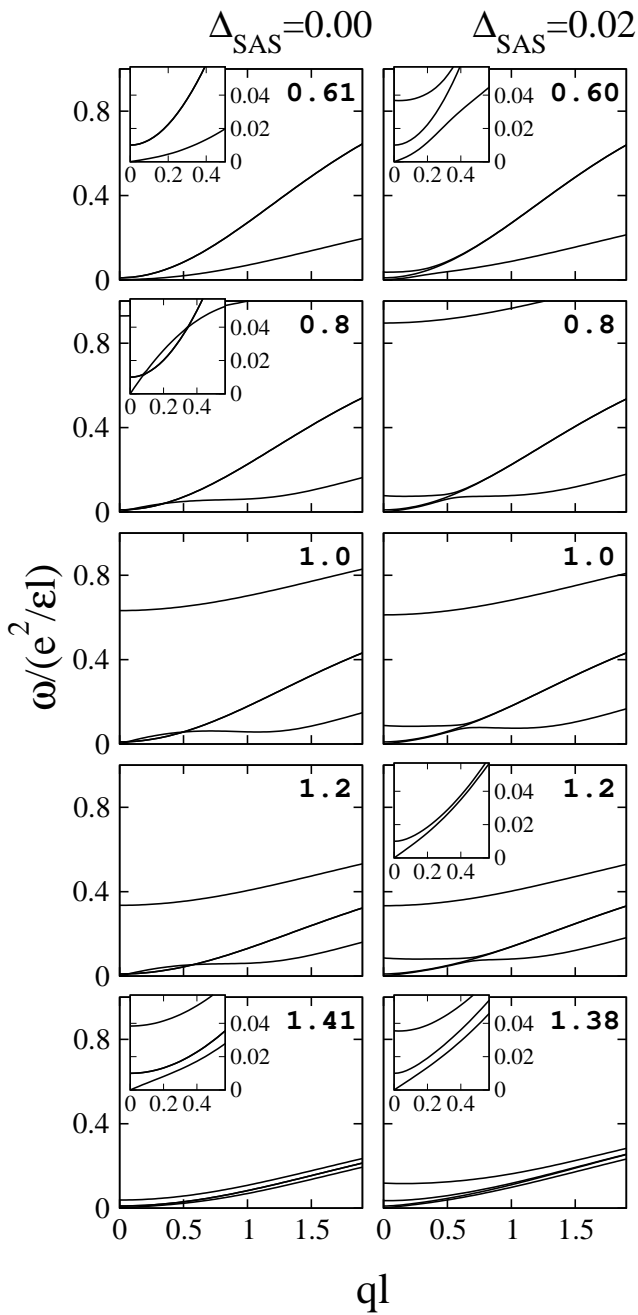}}
    \caption[The collective-mode dispersions of the
    charge-unbalanced $\nu=2$ bilayers in perpendicular field --
    {\it canted and $I$ phases}.]{\label{fig:canted-cmd} The
    collective-mode dispersions of the charge-unbalanced $\nu=2$ bilayers in
     perpendicular field -- {\it many-body phase}. The dispersions in the
     left column are given for a system with $\Delta^0_{SAS} = 0.0(e^2/\varepsilon l)$;
       in the right column -- for a system with
        $\Delta^0_{SAS} = 0.02(e^2/\varepsilon l)$; the
         Zeeman energy in all panels is $\Delta_Z = 0.01 (e^2/\varepsilon l)$; the
          bias voltage is given in the upper right corner of each panel
          in units of $(e^2/\varepsilon l)$.  Some collective-mode
          dispersions are degenerate.  The low-energy, low-wavevector
          region is shown in the insets. }
\end{figure}

\begin{figure}[t]
    \centering
    \scalebox{0.65}{\includegraphics{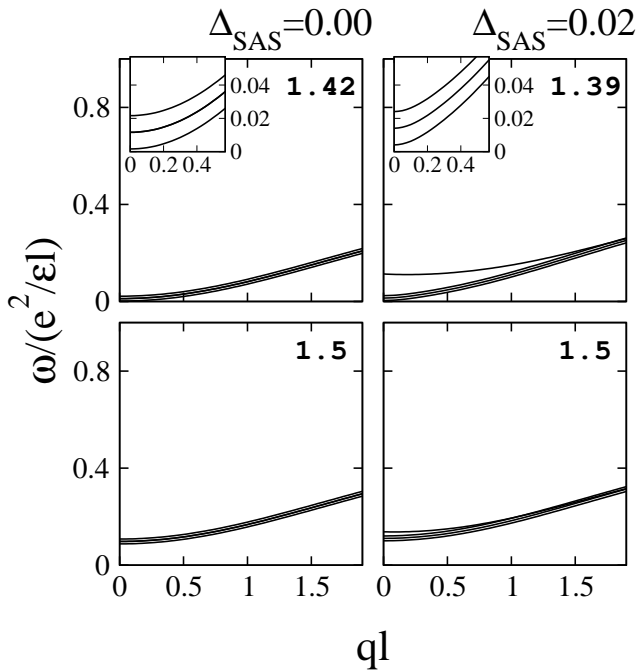}}
    \caption[The collective-mode dispersions of the
    charge-unbalanced $\nu=2$ bilayers in perpendicular field --
    {\it spin-singlet phase}.]{\label{fig:ss-cmd}  The collective-mode
     dispersions of the charge-unbalanced $\nu=2$ bilayers in perpendicular
      field -- {\it spin-singlet phase}. The dispersions in the left column
      are given for a system with $\Delta^0_{SAS} = 0.0(e^2/\varepsilon l)$;
      in the right column -- for a system with  $\Delta^0_{SAS} =
       0.02(e^2/\varepsilon l)$; the Zeeman energy in all panels
       is $\Delta_Z = 0.01 (e^2/\varepsilon l)$; the bias voltage
       is given in the upper right corner of each panel in units
       of $(e^2/\varepsilon l)$.  Some collective-mode dispersions
       are degenerate.  The low-energy, low-wavevector region is shown in the insets. }
\end{figure}

The Hamiltonian of the charge-unbalanced $\nu=2$ bilayer systems
in the presence of interlayer tunneling does not commute with the
$I^z+S^z$ operator.  The symmetry considerations that we used to
obtain an analytical solution for the density-response function of
the $\nu=2$ bilayers in the absence of tunneling cannot be used to
simplify Eq.~(\ref{eq:findpoles}) when tunneling is present.  We
therefore use numerical techniques to calculate the dispersion
relations of the charge-unbalanced $\nu=2$ bilayers with finite
interlayer tunneling.  Much insight into the numerical results can
be gained by comparing the collective-mode dispersions in the
systems with tunneling to the analytical results for the systems
without tunneling.  The comparison is presented in
Figs.~\ref{fig:ferro-cmd}, \ref{fig:canted-cmd}, and
\ref{fig:ss-cmd}, for the ferromagnetic, many-body, and
spin-singlet phases, respectively.  The sets of plots in each
figure are arranged in two columns:  In the left column, we plot
the dispersion curves of a system with no tunneling.  The
dispersions of a system with tunneling are given in the right
column.  In all the plots, the Zeeman energy is set at $\Delta_Z =
0.01 (e^2/\varepsilon l)$;  the external bias voltage is given in
the upper right corner of each panel.

\subsection{Ferromagnetic phase}
\label{subsec:cmd-fer}

In Figure \ref{fig:ferro-cmd}, we present the dispersion curves
obtained in the ferromagnetic phase.  While the ferromagnetic
ground state does not change as the bias voltage is increased, the
collective-mode dispersions demonstrate the evolution of the
inter-subband energetics that eventually leads to a phase
transition.  The top two panels of Fig.~\ref{fig:ferro-cmd} show
the dispersion curves in the absence of bias voltage.  In the
absence of tunneling, we can identify the lower curve as the
degenerate $1\rightarrow 3$ and $2\rightarrow 4$ excitations, that
have a gap equal to the Zeeman energy as $q\rightarrow 0$.  The
Zeeman gap is hard to discern in the figure, since scale of the
Zeeman energy is very small in comparison with the energy scales
of the other excitations.  The upper curve represents the
excitations $1\rightarrow 4$ and $2\rightarrow 3$ that are
degenerate in the absence of tunneling and bias voltage.  When a
small amount of tunneling is present, the degeneracy of the
dispersion curves is lifted, and the curves are split by an energy
of order ${\cal O}(\Delta_{SAS}^2/F_-)$ --- a minute energy
difference, not visible on the scale of the figure.

As the bias voltage is increased, as was discussed in the previous
section, in the absence if tunneling, the splitting between the
dispersions of the $1\rightarrow 4$ and $2\rightarrow 3$ modes
increases linearly with the voltage.  Thus, at $\Delta_V =
0.3(e^2/\varepsilon l)$, the splitting of the upper branches is
very large and apparent.  The Zeeman branch, $1\rightarrow 3$ and
$2\rightarrow 4$, in the systems without tunneling, remains
degenerate for any bias voltage.   Note that, in the absence of
tunneling, the dispersion curve of the $2\rightarrow 3$ mode
crosses the Zeeman branch clearly without interacting with it.
The situation changes when tunneling is present:  the
(approximately) $2\rightarrow 3$ mode develops an anti-crossing
with one of the modes of the Zeeman branch.  The interactions
between different excitations are allowed in the presence of
tunneling by the broken symmetry of the Hamiltonian:  when the
tunneling term is present, the operator $I^z+S^z$ does not commute
with the Hamiltonian; the eigenvalues of the operator are,
therefore, no longer good quantum numbers of the excited states.
Nevertheless, one mode always stays independent of other
excitations:  One can see that, in the states with finite
magnetization, Figs.~\ref{fig:ferro-cmd} and \ref{fig:canted-cmd},
the spin-wave mode, identified by the Larmor minimum at the Zeeman
energy, is always decoupled from the other modes.  This is a
consequence of the up-down, left-right symmetry of the Hamiltonian
mentioned in the Sec.~\ref{subsec:sw} and preserved in the
presence of tunneling.  This symmetry maps the excitation
$1\rightarrow 3$ to $4 \rightarrow 2$.  The mapping results in a
special form of the density-response matrix that always separates
out one Zeeman mode.

\subsection{Many-body canted and $I$-phases}
\label{subsec:cmd-mb}

As the voltage is increased further, the gap of the $2\rightarrow
3$ mode decreases until it disappears.  The softening of the mode
signals the onset of a many-body phase --- the $I$-phase in the
system without tunneling and the canted phase in the presence of
tunneling.   The critical voltage is higher in the system without
tunneling, since, when the tunneling is present, it works in
concert with the bias voltage to stabilize the spin-singlet phase
and destabilize the ferromagnetic phase.  The details of the
collective mode dispersions as the gap of the $2\rightarrow 3$
mode approaches zero around $\Delta_V \approx 0.6(e^2/\varepsilon
l)$ are given in the insets of the bottom two panels of
Fig.~\ref{fig:ferro-cmd}.  In both insets, one can clearly see the
Zeeman branch with the gap at the Zeeman energy, $\Delta_Z =
0.01(e^2/\varepsilon l)$.  In the absence of tunneling, Zeeman
branch is degenerate; when tunneling is finite, the degeneracy is
lifted by the interaction with the $2\rightarrow 3$ branch near $q
= 0$.  The interaction of the $2\rightarrow 3$ excitation with the
superposition of the $1\rightarrow 3$ and $2\rightarrow 4$
excitations result in an earlier onset of the many-body phase and
a mixed mode with a large gap around $q=0$. When the bias voltage
is increased a little above $\Delta_V \approx 0.6(e^2/\varepsilon
l)$, the system undergoes a phase transition into the many-body
phase.  The collective-mode dispersions for the $\nu=2$ bilayers
in the many-body phase are given in Fig.~\ref{fig:canted-cmd}.
Right after the transition the velocity of the Goldstone mode is
very low, but it rapidly increases as the system is taken deeper
into the many-body phase by an increasing bias voltage.  Near the
transition, the velocity of the mode increases faster in the
canted phase, than in the $I$-phase.  In the canted phase,
however, because of the inter-mode interactions, the velocity soon
reaches nearly a constant, while in the $I$-phase it continues
increasing until approximately the middle of the $I$-phase.  As
the system gets closer to the transition to the spin-singlet
phase, the Goldstone-mode velocity goes to 0 in the reverse
fashion.

Another effect of the broken-symmetry of the Hamiltonian in the
presence of tunneling is the further widening of the gap that
develops in the mode that splits off the Zeeman branch as a
consequence of the mixing with the other modes.  This gap is much
larger than $\Delta_{SAS}$ and is therefore due to the
interactions.  Finite $\Delta_{SAS}$ in this case mainly serves to
break the symmetry of the Hamiltonian.  This, ``third'', mode
develops a non-analyticity at $q=0$ and a ring of shallow roton
minima, degenerate for all directions of $q$.

The roton minimum that, as we showed in the previous subsection,
characterizes the $I$-phase is inherited by the low-tunneling
canted phase.  It is however less deep in the canted phase, and
gradually disappears as tunneling is increased until it becomes
more important than the bias voltage (i.e.\@ when the bias voltage
does not result in a significant charge imbalance within the
many-body phase).  Another feature to observe is the lowering of
the energy of the highest-energy mode.  Its energy scale changes
dramatically as one sweeps across the many-body phase by
increasing the bias voltage.  Around the boundary between the
ferromagnetic phase and the many-body phase, the gap of the
highest-energy mode is two orders of magnitude larger than
$\Delta_Z$ (it is rather of ${\cal O}(\Delta_V)$), but at the
boundary of the many-body phases and the spin-singlet phase, the
gap decreases to $2\Delta_Z$.

\subsection{Spin-singlet phase}
\label{subsec:cmd-ss}

When the bias voltage is around $\Delta_V \approx 1.4
(e^2/\varepsilon l)$, the system undergoes a phase transition from
the many-body to the spin-singlet phase.  At this point the
Goldstone branch develops a gap.  The Goldstone mode becomes the
lowest of the three spin-triplet excitations above the
spin-singlet ground state.  As can be seen in the insets, the
three modes are separated by $\Delta_Z$.  In the absence of
tunneling the $S^z = 0$ spin triplet is degenerate with the only
allowed spin-singlet excitation.  When tunneling is present, the
energy of the spin-singlet excitation is affected by the
interactions;  it slowly approaches the energy of the $S^z = 0$
spin triplet as the increasing bias voltage turns the
interlayer-phase coherent spin-singlet state at $\Delta_{SAS} \neq
0$ into a $\nu=2$ spin-unpolarized monolayer state.

% LocalWords:  diagrammatically renormalization DSSZ TDHF ret ig ik xX xk yl iQ
% LocalWords:  CHIS Matsubara KH cabd ab bacd cd subband SAS Shr cccc subbands
% LocalWords:  decoupled gapless RL ql superpositions FERRO pre destabilized nX
% LocalWords:  decouples analyticity energetics Larmor calculational kallin

% LocalWords:  zheng dassarma leaves superfluid superfluid superfluid
% LocalWords:  superfluid superfluids anisotropies

\section{Collective-mode dispersions of charge-unbalanced $\nu=2$ bilayers in tilted field}
\label{sec:cmd-t}

\begin{figure}[t]
    \centering
    \scalebox{0.60}{\includegraphics{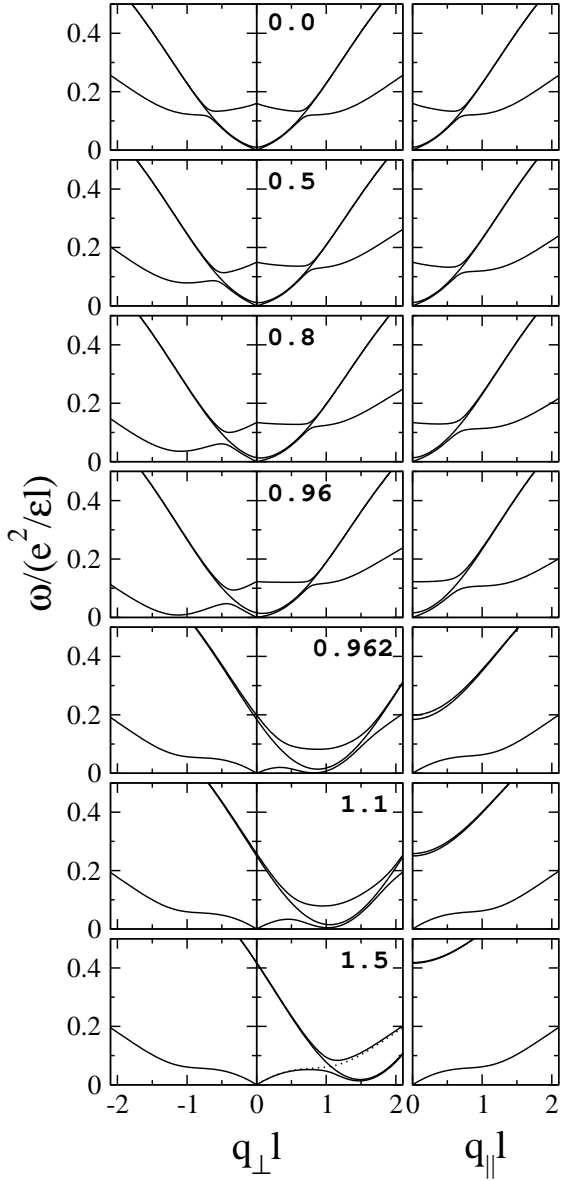}}
    \caption[The collective-mode dispersions of the
    charge-unbalanced $\nu=2$ bilayers in tilted field --
    evolution across the $C1$--$C2$ phase transition.]{\label{fig:c1c2-cmd}
    The collective-mode dispersions of the charge-unbalanced $\nu=2$ bilayers
    in tilted field -- evolution across the $C1$--$C2$ phase transition.  The
    dispersions are given for a system with $\Delta^0_Z = 0.01 (e^2/\varepsilon l)$,
     $\Delta^0_{SAS} = 0.06(e^2/\varepsilon l)$, and $\Delta_V = 0.8 (e^2/\varepsilon l)$.  The strength of the in-plane component of the tilted magnetic
field is given by the wavevector $Q_{||}$, which, for every set of
dispersion curves, is represented by a number in the units of
$1/l$.  When $Q_{||}l \leq 0.96$, the system is in the $C1$ phase;
when $Q_{||}l > 0.96$, it is in the $C2$ phase.  In the bottom
left panel, the dotted line is for comparison of the dispersion
curves in the $C2$ phase to those in the $I$-phase.}
\end{figure}

As we discussed in Sec.~\ref{sec:unres} above and in
Ref.~\onlinecite{lopatnikova:03a}, tilting the magnetic field away
from the normal to the plane of the bilayer system leads to new
phases and, in charge-unbalanced systems, phase transitions. In
the charge-balanced $\nu=2$ bilayers, the tilted fields do not
change the topology of the phase diagram. The interlayer phase
coherent phases -- spin-singlet and canted -- become commensurate,
and the ferromagnetic state is not affected by the in-plane field.
In commensurate interlayer phase coherent phases, the interlayer
exchange interactions are effectively destabilized by the in-plane
field, and the phase-space volume of the canted phase decreases as
the magnetic field is tilted.  The weakening of interlayer
exchange interactions renormalizes collective mode dispersions of
the charge-balanced $\nu=2$ bilayers but does not result in
interesting new features.

The presence of a finite in-plane component of the magnetic field
produces more interesting effects in charge-unbalanced systems.
When tunneling is strong enough, the in-plane field induces a new
phase transition between the simple commensurate phase, $C1$,
stable at low in-plane fields, and the spin-isospin commensurate
phase, $C2$, more favorable at higher in-plane fields.  As is
shown in Fig.~ \ref{fig:c1c2single} and discussed in depth in Ref.
\onlinecite{lopatnikova:03a}, the phases $C1$ and $C2$ are
connected to each other, and the first-order transition between
them terminates at a critical point.  To further study this
first-order transition, we obtain a series of collective-mode
dispersions, calculated for the $C1$ and $C2$ states as the
in-plane field is increased.  We also obtain the collective-mode
dispersions at the critical point terminating the $C1$--$C2$
transition.

\subsection{Collective-mode dispersions across the $C1$--$C2$ phase transition}
\label{subsec:cmd-cc}

The evolution of the dispersion curves within the canted phases as
the magnetic field is tilted is given in Fig.~\ref{fig:c1c2-cmd}.
We choose a system with $\Delta^0_Z = 0.01 (e^2/\varepsilon l)$
and $\Delta^0_{SAS} = 0.06 (e^2/\varepsilon l)$, and hold the
external bias voltage at $\Delta_V = 0.8 (e^2/\varepsilon l)$, so
that the system is approximately in the middle of the canted phase
(c.f.~Figs.~\ref{fig:c1c2single}).  For each probed point on the
phase diagram in Fig.~\ref{fig:c1c2single}, we plot in
Fig.~\ref{fig:c1c2-cmd} the cross-sections of the collective-mode
dispersions in two directions -- perpendicularly to the in-plane
field (in the $\hat{x}$ direction in our calculations), and in the
direction parallel to it (the $\hat{y}$ direction);  these plots
are given side by side.  The measure of the magnitude of the
in-plane component of the magnetic field, the wavevector $Q_{||}$,
is given as the number in the left panel.

The top two panels show the collective-mode dispersions in
perpendicular magnetic field.  In perpendicular field, the
dispersions are the same in all directions, so the dispersion
curves in the left and the right panels coincide.  One can see the
features discussed for the canted phase in the charge-unbalanced
$\nu=2$ bilayers (Sec.~\ref{sec:cmd-00}):  the linearly dispersing
Goldstone mode that has a roton minimum around $ql \approx 1$,
characteristic of charge-unbalanced systems;  the spin-wave mode
that decouples from the other three modes and has a gap equal to
the Zeeman energy (the resolution of the figure does not allow to
see the Zeeman splitting because of the relatively small Zeeman
energy);  the large interaction-induced gap at $q=0$ of the third
mode.  The highest-energy mode is not visible in this figure.

When the magnetic field is tilted, the collective modes start
changing:  they become asymmetric with respect to $q_{\perp}
\rightarrow -q_{\perp}$.  The velocity of the Goldstone-mode in
the negative $q_{\perp}$-direction becomes greater than that in
the positive $q_{\perp}$-direction.  The roton minima also become
asymmetric --- they develop a lowest point in the negative
$q_{\perp}$-direction.  This behavior is reminiscent of the
behavior of the collective-mode dispersions of a (charged)
superfluid under the influence of an external electromagnetic
gauge field, ${\bf A}$.  In a superfluid, much like in the $\nu=2$
bilayers in the canted phase, a U(1) symmetry is spontaneously
broken.  The symmetry breaking results in the formation of a
linearly dispersing Goldstone mode.  When an external field is
applied to a charged superfluid, the superfluid order parameter
acquires a twist, $e^{i \frac{e}{c}{\bf A} \cdot {\bf x}}$, and
the Goldstone mode acquires an anisotropy: \be
    \omega_{\bf k} =  |{\bf k}| \sqrt{v_0^2 - \frac{1}{2}(\frac{\hbar e}{m c})^2 |{\bf A}| ^2} - \frac{\hbar e}{m c} {\bf A}\cdot {\bf k}, \label{eq:doppler}
\ee where $v_0$ is the initial velocity of the Goldstone mode.
While the in-plane field does not couple to the symmetry-breaking
order parameters in $\nu=2$ bilayers in the same way it does in a
superfluid, it does result in winding phase factors $e^{i Q_{S}
X}$, $e^{i Q_{I} X}$, and $e^{i (Q_{S}\pm Q_{I}) X}$.  Because of
the gauge symmetry of our system, an equivalent picture can be
drawn up, in which fictitious gauge fields proportional to
$Q_{S}$, $Q_{I}$, and $Q_{S}\pm Q_{I}$ couple to the corresponding
(uniform) order parameters.  In fact, this is exactly what we have
done, when we chose to work in the basis of the
creation-annihilation operators $f_{\alpha}$ (see
Sec.~\ref{sec:drf} and Eq.~(\ref{eq:rhof})), in terms of which the
ground state is uniform.

Because the in-plane field generates different phase factors
($e^{i Q_{S} X}$, $e^{i Q_{I} X}$, and $e^{i (Q_{S}\pm Q_{I}) X}$)
for different order parameters, the $\nu=2$ bilayer system is
somewhat more complicated than a model superfluid.  It is clear in
Fig.~\ref{fig:c1c2-cmd}, that different collective-mode branches
(and different parts of the branches) do not respond to the
presence of an in-plane field in the same way.  Thus, the
Goldstone mode ``tilts'' to the right in Fig.~\ref{fig:c1c2-cmd},
while the roton minima ``tilt'' to the left.  An effective theory
of coupled superfluids that would explain the behavior of the
collective modes, as well as the $C1$--$C2$ superfluid-superfluid
phase transition is a potential direction of future research.

In addition to the anisotropies, the application of the in-plane
field results in an increased Zeeman mode gap, $\Delta_Z =
\Delta^0_{Z}\sqrt{1+Q_{||}^2l^4/d^2}$.  It is also apparent,
especially at higher in-plane field, that the minimum of the
Zeeman mode shifts from $q_{\perp}=0$ to $q_{\perp}=Q_S$.  This
effect is also a consequence of the our choice of the gauge.  As
was mentioned in Sec.~\ref{sec:drf}, the density response function
$\chi_{\alpha\beta\gamma\delta}({\bf k};\tau)$ that we calculate
is gauge-dependent, because it is the response of the system to
the excitation with the density operator  $\rho_{\alpha\beta}({\bf
k})$.  This density operator is related to the physical density
operators not only through a linear transformation, but also
through a shift of the wavevector (Eq.~(\ref{eq:rhoab})).  A real
excitation therefore will pick out signals from different
dispersion curve branches (according to allowed symmetries) at
different wavevectors.  Thus, for example, the physical response
function for a real spin-flip excitation of wavevector ${\bf k}$
is a superposition of the $\chi_{\alpha\beta\gamma\delta}({\bf
q};\tau)$ at wavevector ${\bf k}+Q_S\hat{x}$ \bea
    \lefteqn{\chi_{\mu\downarrow\,\mu\uparrow ;\mu\downarrow\,\mu\uparrow}
    ({\bf k},\tau) = -g\langle T\rho_{\mu\downarrow\,\mu\uparrow}
    ({\bf k},\tau)\rho^{\dag}_{\mu\downarrow\,\mu\uparrow}({\bf k},0)\rangle = }
    \label{eq:chishift}\\
    && = e^{i\mu Q_{||}k_yl^2} \sum_{\alpha\beta\gamma\delta}
    (z^{\alpha}_{\mu\downarrow})^*z^{\beta}_{\mu\uparrow}
    (z^{\delta}_{\mu\uparrow})^*z^{\gamma}_{\mu\downarrow}\,
    \chi_{\alpha\beta\gamma\delta}({\bf k}+Q_S\hat{x},\tau)). \nonumber
\eea

The dispersion curves in the direction parallel to the in-plane
field stay symmetric with respect to $q_{||}\rightarrow -q_{||}$.
As the tilt-angle is increased, they change mainly because the
minimum of the Zeeman and other branches shift away from the
$q_{\perp} = 0$ plane plotted in the right column of
Fig.~\ref{fig:c1c2-cmd}.

As the magnetic field is tilted further, the anisotropy of the
Goldstone-mode velocity becomes greater, and the energy of its
roton minimum around $q_{\perp}l \approx -1$ decreases.  Near the
$C1$--$C2$ phase transition, $Q_{||} = 0.96$, the roton minimum
becomes lower than the Zeeman energy.  However, before it reaches
zero and the system becomes unstable, the phase transition to the
$C2$ phase occurs, marked by an abrupt change in the
collective-mode dispersions.  The entire picture is effectively
shifted by $Q_{S_{C2}}-Q_{S_{C1}}$ (see Fig.~\ref{fig:c1c2-cmd})
in the positive $q_{\perp}$ direction.  A Goldstone mode appears
in place of the roton minimum, and a roton-minimum replaces the
Goldstone mode.  The minimum of the Zeeman mode jumps from
$Q_{S_{C1}}$ to $Q_{S_{C2}}$, as expected from
Eq.~(\ref{eq:chishift}) and Fig.~\ref{fig:c1c2-cmd}.

\begin{figure*}[t!]
    \centering
    \scalebox{0.7}{\includegraphics{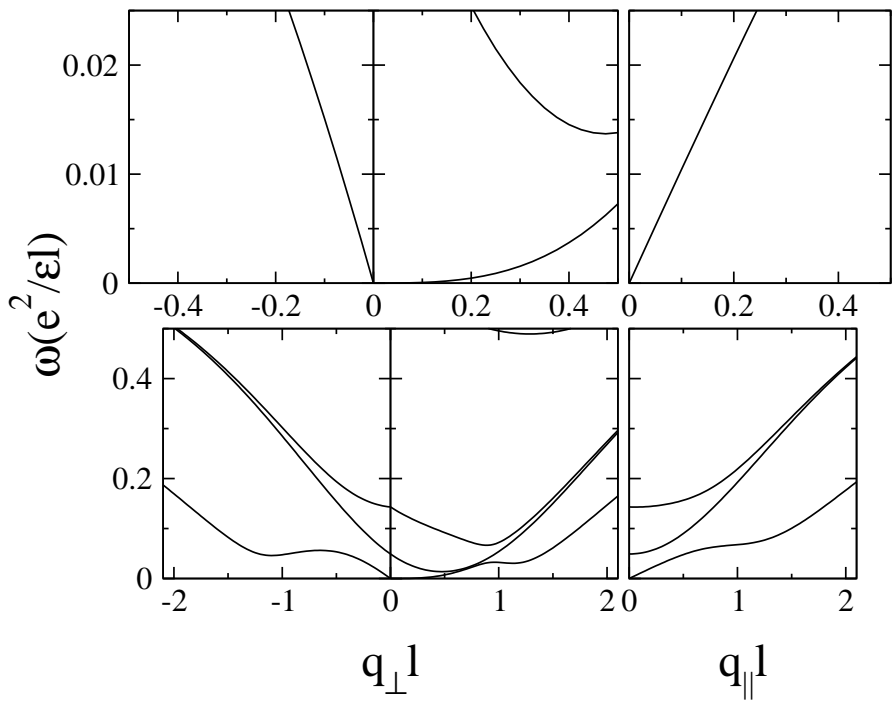}}
    \caption[The collective-mode dispersions of the
    charge-unbalanced $\nu=2$ bilayers in tilted field at a
critical point.]{\label{fig:cep-cmd}  The collective-mode
dispersions of the charge-unbalanced $\nu=2$ bilayers in tilted
field at a critical point terminating the $C1$--$C2$ transition.
The dispersions are given for a system with $\Delta^0_Z = 0.01
(e^2/\varepsilon l)$ and $\Delta^0_{SAS} = 0.06(e^2/\varepsilon
l)$.  The critical point for this system occurs at $\Delta_V =
0.5328$ and $Q_{||} = 0.9362$; $Q_S \approx 0.474$ for these
parameters. The top plot zooms in on the long-wavelength region,
to demonstrate the $q^3_{\perp}$ dependence of the Goldstone-mode
dispersions in the positive $q_{\perp}$ direction. }
\end{figure*}

As the in-plane field is increased further, the roton minimum of
the Goldstone branch becomes less deep -- it approaches the Zeeman
energy, $\Delta_Z$, from below.  When the in-plane component of
the magnetic field becomes large, $Q_{||} \approx 2$, the $C2$
phase becomes very close to an $I$-phase.  This is reflected in
the collective mode dispersions.  In the bottom panels of
Fig.~\ref{fig:c1c2-cmd}, at $Q_{||}$ we can see the symmetry of
the Goldstone branch reappearing (the dotted line in
Fig.~\ref{fig:c1c2-cmd} is given as a guide to the eye).  One of
the Zeeman branches of the $I$-phase forms an anti-crossing with
the Goldstone branch at $q_{\perp} \approx 1$.  When $Q_{||}
\approx 2$ the Zeeman modes and the Goldstone branch become nearly
decoupled.  The Zeeman branches in our calculation have a minimum
at $Q_S \approx Q_{||}$ in high magnetic fields.  This is again
the gauge effect we described above.  Zeeman modes shifted by
$-Q_S$, as they would be in terms of physical densities, would
restore the $I$-phase-like appearance of the collective-mode
dispersions deep in the $C2$ phase.

\subsection{Collective-mode dispersions at the critical point}
\label{subsec:cmd-cep}

In the last part of this section, we consider the collective mode
dispersions at the critical point.  Again, we use a bilayer system
with  $\Delta^0_Z = 0.01 (e^2/\varepsilon l)$ and $\Delta^0_{SAS}
= 0.06 (e^2/\varepsilon l)$.  The critical point for this sample
occurs at $\Delta_V = 0.5328$ and $Q_{||} = 0.9362$; $Q_S \approx
0.474$ for these parameters.  The spin-wave wavevector $Q_S$ is
hard to define precisely at the critical point, since the energy
profile as a function of $Q_S$ is very flat: $~ (Q_S-Q_{S_c})^4$.
The ``flatness'' of the energy as a function of the spin-wave
wavevector $Q_S$ implies the existence of very soft spin-wave
fluctuations.  Indeed, as shown in Fig.~\ref{fig:cep-cmd}, the
velocity of the Goldstone mode in the positive $q_{\perp}$
direction becomes 0 (and the next-order in $q_{\perp}$ emerges:
$\omega \propto q_{\perp}^3$).

% LocalWords:  diagrammatically renormalization DSSZ TDHF ret ig ik xX xk yl iQ
% LocalWords:  CHIS Matsubara KH cabd ab bacd cd subband SAS Shr cccc subbands
% LocalWords:  decoupled gapless RL ql superpositions FERRO pre destabilized nX
% LocalWords:  decouples analyticity energetics Larmor calculational kallin

% LocalWords:  zheng dassarma leaves superfluid superfluid superfluid
% LocalWords:  superfluid superfluids anisotropies

\section{Conclusions}
\label{sec:concl4}

In summary, we have obtained the collective mode dispersions of
the charge-unbalanced $\nu=2$ bilayers in tilted magnetic fields
using the time-dependent Hartree-Fock approximation
(Figs.~\ref{fig:ferro-cmd}--\ref{fig:cep-cmd}).  The collective
modes possess a number of characteristics that can be observed in
light scattering experiments --- such as softening modes and roton
minima.  Thus, the novel phase $C1$--$C2$ phase transition,
discussed at length in Ref.\onlinecite{lopatnikova:03a}, is
signaled by a near-softening of a roton minimum.

The collective-mode dispersions of the $\nu=2$ bilayers in tilted
fields exhibit behavior suggestive of a system of coupled
superfluids under the influence of an external gauge field.  Thus,
when the magnetic field is tilted, i.e.\ a finite in-plane
magnetic field is added to the system, the modes become
Doppler-shifted (Eq.~(\ref{eq:doppler})).  The Doppler shift is
varies for different modes, reflecting on the fact that the
in-plane magnetic field couples to the order parameters of the
$\nu=2$ bilayer systems in different ways (see
Sec.~\ref{subsec:cmd-cc}).

An interesting future direction of research therefore would be the
construction of an effective model with two order parameters that
spontaneously break $U(1)$ symmetry.  An external gauge field can
couple to the order parameters differently, so that when the
breakdown of one superfluid occurs, the other superfluid is
stable.  Such a superfluid transition would be an interesting
model for the $C1$--$C2$ transition in the $\nu=2$ bilayers.
%
%In fact, the author of this thesis, in collaboration with A.~Nihat
%Berker, has studied a similar transition in the ${}^3$He--${}^4$He
%mixtures in silica aerogel \cite{lopatnikova:97a,lopatnikova:97b}.
%In this system, superfluidity is stabilized near the aerogel
%strands and is approximately unaffected by the aerogel in the
%bulk.  As the chemical potential of the system is increased, the
%onset of superfluidity therefore occurs first around the connected
%and spatially continuous aerogel structure;  a first-order phase
%transition into bulk superfluidity follows.  This
%superfluid-superfluid phase transition, not surprisingly,
%terminates at a critical point
%\cite{lopatnikova:97a,lopatnikova:97b}.  It would be interesting
%to find the (universal) critical exponents of the two-superfluid
%model that will describe both $C1$--$C2$ phase transition in the
%$\nu=2$ quantum Hall bilayers and the superfluid-superfluid
%transition of the ${}^3$He--${}^4$He mixtures.

\section*{Acknowledgements}

The authors would like to acknowledge helpful conversations with
J.~H.~Cremers, S.~Das Sarma, B.~I.~Halperin, L.~Marinelli,
A.~Pinczuk, D.~Podolsky, L.~Radzihovsky, G.~Refael,
Y.~Tserkovnyak, D.-W.~Wang, and X.-G.~Wen.  This work has been in
part supported by NSF-MRSEC grant No.~DMR-02-13282 and by the NSF
grant No.~DMR-01-32874.  A.L. would like to thank Lucent
Technologies Bell Labs for hospitality and support under the GRPW
program.

\end{document}